\documentclass[nohyper,notoc]{article} 
\usepackage{axodraw}
\usepackage{epsfig}
\usepackage{bm}
\usepackage{color}
\usepackage[dvipsnames]{xcolor}
\usepackage{amsmath}

\usepackage[utf8]{inputenc} 
\usepackage[T1]{fontenc}    

\def\ben{\begin{enumerate}}
\def\een{\end{enumerate}}
\def\bit{\begin{itemize}}
\def\eit{\end{itemize}}
\def\beq{\begin{equation}}
\def\eeq{\end{equation}}
\def\bea{\begin{eqnarray}}
\def\eea{\end{eqnarray}}
\def\bq{\begin{quote}}
\def\eq{\end{quote}}
\def \lsim{\mathrel{\vcenter
     {\hbox{$<$}\nointerlineskip\hbox{$\sim$}}}}
\def \gsim{\mathrel{\vcenter
     {\hbox{$>$}\nointerlineskip\hbox{$\sim$}}}}
\def\gappeq{\mathrel{\rlap {\raise.5ex\hbox{$>$}}
{\lower.5ex\hbox{$\sim$}}}}
\def\lappeq{\mathrel{\rlap{\raise.5ex\hbox{$<$}}
{\lower.5ex\hbox{$\sim$}}}}

\def\LNP{\Lambda_{\rm NP}}
\def\meg{\mu \to e \gamma}

\def\muc{\mu A\to e A}
\def\mec{\mu \! \to \! e~ {\rm conversion}}

\def\meee{\mu \to e \bar{e} e}

\def\a{\alpha}
\def\b{\beta}
\def\g{\gamma}

\begin{document}

\renewcommand{\thefootnote}{\fnsymbol{footnote}}
\begin{center}
{\Large {\bf
What  is Leading Order for LFV in SMEFT?
}}
\vskip 20pt
{\large Marco Ardu \footnote{E-mail marco.ardu@umontpellier.fr},
and 
Sacha Davidson \footnote{E-mail address:
    s.davidson@lupm.in2p3.fr} }
 
 \vskip 10pt  

{\it LUPM, CNRS,
Université Montpellier
Place Eugene Bataillon, F-34095 Montpellier, Cedex 5, France
}\\

\end{center}
\begin{abstract}
\noindent
Upcoming  searches for lepton flavour change (LFV) aim to probe
New Physics(NP) scales up to $\Lambda_{NP} \sim 10^4$ TeV, implying
that  they will be sensitive to NP at lower scales that is suppressed
by loops or  small couplings. 
We suppose that the NP responsable for
LFV  is beyond the reach of the LHC  and
can  be parametrised in Effective Field Theory,
introduce a small power-counting
parameter $\lambda$ (\`a la Cabibbo-Wolfenstein), and  assess whether
the existing dimension six operator basis and one-loop RGEs
provide a good approximation for LFV.  We 
find that $\mu \leftrightarrow e$ observables  can be sensitive to
a few dozen dimension eight operators,  and to some effects of two-loop
anomalous dimensions, for  $\Lambda_{NP} \lsim 20-100$ TeV.  We also
explore the effect of some simplifying assumptions in the one-loop RGEs,
such as  neglecting flavour-changing effects.

\end{abstract}

\section{Introduction}
\label{intro}

Perturbation theory is a  widely used tool in the Standard
Model (SM), New Physics (NP) models and many other  areas.
In a given perturbative expansion, the first  non-vanishing
term,  sometimes called the  leading order contribution,
is often simple to compute. However,  when a calculation
simultaneously involves many perturbative expansions,
it can be more challenging  to identify the  ``leading'' or dominant
contribution.

In this manuscript, we  study perturbative expansions
in the Lepton Flavour Changing(LFV)  part of
the  Lagrangian \cite{BW,polonais}
of the Standard Model Effective Field Theory (SMEFT). 
We restrict to LFV operators  for two reasons ; firstly,
they  must exist because the observations of neutrino
oscillations  demonstrate that leptons change flavour
(for a
review of LFV in the $\mu\leftrightarrow e$ sector, see \cite{KO}).
And secondly, LFV operators are simpler than generic operators, because
SM  loop effects, included via Renormalisation Group
Equations(RGEs),  cannot  change lepton flavour, so 
the flavour of at least two legs of each operator remains fixed.

There are many perturbative expansions in SMEFT:
the EFT expansion in the ratio of weak to New Physics scales $v^2/\Lambda_{NP}^2$, as well as
the SM  expansions  in loops, in the ${\cal O}(1)$
gauge and Higgs self-couplings and   in the exceptionally hierarchical
Yukawa couplings, and also in mixing angles. 
So it is not obvious to find the  leading effects.
For example, it was noticed long ago by Bjorken and
Weinberg\cite{BjW}, in the SM extended with a  second Higgs
$H$ with LFV couplings $Y_{\mu e}\bar{\ell}_\mu H P_R e $,  that
the one-loop amplitude for $\meg$ is  suppressed by
two lepton Yukawas, so  is {\it smaller} than two-loop
(``Barr-Zee''\cite{BZ}) contribution:
$$
{\cal A}_{1-loop} \propto\frac{ey_\mu Y^*_{\mu \mu} Y_{\mu e}  }{16\pi^2 M_H^2}
~~~,~~~
{\cal A}_{2-loop} \propto\frac{ey_t g^3 Y_{\mu e}  }{(16\pi^2)^2 M_H^2} ~~~.
$$
However, this leading  (although two-loop)  contribution
was missed in part of the subsequent literature.

In the QCD$\times$QED-invariant EFT below the weak scale,
various powercounting schemes  exist to organise
perturbative calculations. For instance in the quark flavour
sector, the  Wolfenstein parametrisation
of the CKM matrix \cite{W}
 in powers of Cabibbo's $\lambda \sim 0.22$, allows
to guess the  order of diagrams\cite{burashouches,Luca}. And in
the RGEs for four-quark operators,  QED
effects can be included at  appropriate subleading
order in the   expansion in $\a_s \log$\cite{burashouches,Huber:2005ig}.
For LFV below the weak scale,
 the ``leading order''
operators and RGEs  have been assembled:
observables
can be parametrised with
three and four-point functions, which correspond to operators  of
dimensions five to eight (see {\it eg} \cite{C+C} for  a list),
and the ``leading order'' RGEs, which
include  two-loop vector to dipole mixing,
are given in \cite{PSI}. 
However, above the weak scale, the situation is complicated  by
the dynamical Higgs and SU(2) gauge bosons, which
introduce more particle mixing in
the  RGEs,  
 and also
by our ignorance of the  mass scale of
new particles $\LNP$.

In this manuscript, we suppose
that New Physics is  ``beyond the  LHC'',
which is taken to mean $\Lambda_{NP}> 4 $ TeV,
and  introduce in Section \ref{sec:powercount} 
a generalisation  of the  Wolfenstein  counting
\footnote{We thank Junji Hisano for proposing
  the original scheme.}
that parametrises the expansions in  all the SM parameters of SMEFT, as
well as the scale ratio $v/\LNP$, in terms of a  single power-counting parameter
$\lambda\sim 0.2$. 
For any operator, this scheme allows to
identify the ``leading''
contribution to a given process  among those that
 could  arise at different orders in the  multiple perturbative expansions.
 It also allows to classify the contributions
of various operators to a process according to the
order in $\lambda$, and  estimate when a
process can have sensitivity to an operator.
So in section \ref{sec:qestions},
 the power-counting scheme is used to address four questions:
\ben
\item  Are dimension six operators sufficient to parametrise
  LFV,  or  can observables be sensitive to
  dimension eight operators?
\item Does one need two-loop   anomalous dimensions in  the  RGEs?
\item  Are LFV observables sensitive to the effects
  of CKM angles in the RGEs, or can the quark Yukawa
  matrices be approximated as diagonal?
\item If the dimension six operator $H^\dagger H \overline{\ell} H e $
is present, it contributes to the charged lepton mass matrix
when the Higgs has a vev, so  the lepton mass eigenstates
are not  the eigenstates of the  lepton Yukawa  matrix
$Y_e$ that appears in the RGEs. How  should this be accounted for?
\een

The results are summarised in section \ref{sec:summ}. The powercounting
suggests that  in the $\mu\leftrightarrow e$ sector,
upcoming data could be sensitive to some dimension eight operators,
and  some ${\cal O} (\log/(16\pi^2)^2)$ effects,
for  4 TeV $\lsim \LNP \lsim$
100 TeV. The relevant dimension eight
operators are listed in  Appendix \ref{app:O8},
and their (tree-level) matching onto the
EFT below $m_W$ is given in Appendix \ref{app:match}.
For $\LNP \gsim 50\to 100$ TeV
in the  $\mu\leftrightarrow e$ sector, and   for
all considered scales 
in the  $\tau\leftrightarrow \ell$ sector ($\LNP \gsim 4$ TeV),
the powercounting 
suggests that the one-loop RGEs for dimension six operators are sufficient.

\section{Power-counting}
\label{sec:powercount}

We want to connect  low-energy LFV processes
 with  the operator coefficients  in the SMEFT.  In a top-down
sense, this
means we want  to estimate the ``leading'' or largest  contribution 
of each operator coefficient to each observable, or equivalently from a
bottom-up perspective,
 the best sensitivity of each observable to each operator.

\subsection{Notation}

We write   the  SM Lagrangian  in notation
similar to  \cite{JMT};
the differences are  that the index order on Yukawas is doublet-singlet,
and $\lambda$ will be the small  power counting parameter,
rather than  the   Higgs self-coupling.

Operators that change lepton flavour (but not
number) 
  arise at
dimension $\geq 6$ in SMEFT, and are added to the Lagrangian
in the basis  of \cite{BW, polonais},
with  coefficients  written as a  dimensionless  $C$
 divided by appropriate factors of a mass  scale  $\Lambda $:
\beq
{\cal L}_{SMEFT} = {\cal L}_{SM} +
{\Big \{}
\frac{1}{\Lambda^2}
\sum_I C_I^{\zeta}{\cal O}^{\zeta}_I
+ \frac{1}{\Lambda^4} \sum_K ~^{(8)}C_K^{~\zeta}{\cal O}^{(8)~\zeta}_K+ h.c.   +...{\Big \}}
\label{L} 
\eeq
where $\Lambda$
is   $v = 174$ GeV in the experimental
constraints on coefficients ($1/v^2 = 2\sqrt{2} G_F$),
but 
it is sometimes convenient
in the powercounting to take $\Lambda$ to be
the  scale  $\LNP$ of New LFV Physics.
So  for $\Lambda = v$,
the powers of $v^2/\LNP^2$ are   included  the coefficients
$C$. The coefficient   subscripts label
the gauge structure, and the superscript $\zeta$
is the flavour of the fermions  composing the operator
in order of appearance(sometimes
the LFV indices are suppressed when they are obvious). 
The dimension six  and eight operators  are
respectively labelled and  normalised
as in \cite{polonais}\footnote{The hermitian operators
are here defined with a 1/2, since the hermitian conjugates
are included in eqn \ref{L}.} and \cite{Murphy:2020rsh}.

The operators  $\{{\cal O}_I\}$ represent LFV contact interactions
among SM particles.  Loop corrections to the operators generically diverge,
so after renormalisation in $\overline{MS}$, the  operator
coefficients  depend on the renormalisation
scale $\mu$ and satisfy Renormalisation Group Equations (RGEs).
These can be written  for dimension six operators as
\bea
\mu \frac{\partial }{\partial \mu}\vec{C} = \frac{1}{16\pi^2}
\vec{C} \widetilde{\Gamma}  + ...
\label{RGE}
\eea
where the operator coefficients are lined up in the row
vector $\vec{C}$, and the matrix elements of  $ \widetilde{\Gamma}$
are the  anomalous dimensions multiplied by SM couplings,
currently known at one-loop (see $eg$ \cite{JMT}).
The matrix $\widetilde{\Gamma} $ can be improved by including
higher-loop contributions to the anomalous dimensions, and
the  equation can be extended by adding higher-dimensional
operators (which changes its structure\cite{Herrlich:1994kh}).
Eqn(\ref{RGE}) can be solved numerically, or solved analytically  as a 
``scale-ordered'' exponential, or approximated
by neglecting the running of SM couplings
and exponentiating $\widetilde{\Gamma}$:
\beq
\vec{C}(\mu_{2}) \simeq \vec{C}(\mu_{1}) +
\vec{C}(\mu_{1})  \frac{\widetilde{\Gamma}}{16\pi^2} \ln\left(\frac{\mu_{2}}{\mu_{1}}\right)+... 
\label{solnRGEs}
\eeq
 This last  approximation can be improved
 by including the running of some SM  couplings,
 and selected  ${\cal O}(\ln^2/(16\pi^2)^2)$
 terms. The power-counting scheme  introduced below
 is diagrammatic, so  makes estimates in the
 spirit of  an improved eqn (\ref{solnRGEs}), and 
 aims  to assist in determining which
 improvements should be included in the RGEs.

\subsection{The power-counting scheme}
\label{ssec:lambda}

The aim here is to construct  a power-counting scheme
allowing to organise 
the perturbative expansions that
arise in  Renormalisation Group running
in the SMEFT above $m_W$.
The input to this  power-counting scheme  should be 
the experimental sensitivities   of one or several observables,
and a list of   operator coefficients.  But since one
of the  expansion parameters, $v^2/\LNP^2$, is unknown,
we ony bound it from above,
and quantify the order of a coefficients contribution
to an observable,  as the scale
up to which an ${\cal O}(1)$ coefficient could  be probed.

We introduce a small parameter 
\beq
\lambda \simeq 0.2
\label{lambda}
\eeq
by analogy  to the $\lambda$ parameter of the CKM matrix.  The numerical value
of  powers $\lambda^k$ is given in table \ref{tab:0}. The various
dimensionless
expansion parameters that occur in SMEFT  can be associated
to powers of $\lambda$  as discussed below (the recipe is
summarised in table \ref{tab:pc}).

\begin{table}[ht]
$\!\!\!\!\!\!\!\!\!\!\!\!$
\begin{tabular}{|c|c|c|c|c|c|c|c|c|c|c|}
\hline $k=$&1&2&3&4&5&6&7&8&10&12\\
\hline
$\lambda^k$ =&.2&.04&.008&
$\!1.6\times 10^{-3}\!$&$\!3.2\times 10^{-4}\!$&$\!6.4\times 10^{-5}\!$&
$\!1.28\times 10^{-5}\!$&$\!2.56\times 10^{-6}\!$
&$\!1.02\times 10^{-7}\!$& $4\times 10^{-9}$\\ \hline
$\LNP$(TeV)&&&&4.3 &&22&&109 & 540&2700\\ \hline 
\end{tabular}
\caption{The second line gives the numerical
value of $\lambda^k$, for $\lambda = 0.2$ and $k$ from the
first line. The third line gives
the value of $\LNP$, in TeV,  such that $(v/\LNP)^2 = \lambda^k$ (where
$v=174$ GeV).
\label{tab:0}}
\end{table}

\ben

\item 
the gauge couplings
$g_s,g_2$ and $g'$ (of respectively QCD, SU(2) and
hypercharge) are counted $\sim {\cal O}(1)$,
and  sometimes retained
in the estimates (because $e^3 \sim \lambda^2$).

\item   With a Lagrangian normalised
  as eqn (\ref{L}) with   $\Lambda = v =174$ GeV,
the  ratio $v^{4-n}/\LNP^{4-n}$ is absorbed into the  coefficients.
In discussing dimension eight
  operators, we assume
  a New Physics scale beyond the reach of the LHC:  
  $$\LNP\gsim  4 ~{\rm  TeV} ~~~~\Rightarrow \frac{v^2}{\LNP^2} \lsim \lambda^4
$$
however  we leave $\LNP>v$ undetermined  
  in estimating the relevance of two-loop or CKM  effects.

\item to each  loop   is attributed a factor
$$\frac{1}{16 \pi^2 } \sim \lambda^3 ~~~,~~~
\frac{ \log}{16 \pi^2 }  \sim \lambda^2 ~$$
where the loops that appear in the RGEs are accompanied by a
 log,  so  counted with one less power.
 (For reference,  $\ln\frac{m_W}{m_{\mu,\tau}}\simeq 6.7, 3.85 $,
and $\ln\frac{4 {\rm TeV}}{m_W}\simeq 3.91$.)

\item anomalous dimensions are counted as ${\cal O}(1)$,
despite that   some  can be large 
(this may sometimes  compensate for  counting
gauge couplings $\sim 1$).

\item In the lepton sector, we  work in the
mass eigenstate basis for charged leptons.
This would be the  eigenbasis of  $Y_e$
in the SM,  but  can differ in
the presence of  non-renormalisable   operators\cite{GL}.
For instance, the operator  
 $ [C_{eH}]^{ij}/\LNP^2 ~ H^\dagger H\bar{\ell}_i H e_j$
 contributes to the  charged lepton mass matrix 
 \beq
[m_e]^{ij} = [Y_e]^{ij}v - [C_{eH}]^{ij}\frac{v^3}{\LNP^2}~~.
\label{me}
\eeq
However, there is a 3 in the Feynman rule of
${\cal O}_{eH}$, such that the  coupling of
leptons to the SM Higgs is 
\beq
    [\widetilde{Y}]_{ij} 
    = \frac{1}{\sqrt{2} v}
    \left([m]_{ij} -  2C^{ij}\frac{v^3}{\Lambda_{NP}^2}\right)
\label{appFR:Y1}
\eeq
so in the charged lepton mass basis, flavour-changing higgs
decays probe the off-diagonal coefficients of $C_{eH}$.

The LHC measures  the yukawas
  of  the  $\tau$ and the  $\mu$ to be consistent with 
  SM  expectations \cite{CMStau,CMSmuon},
  and constrains the $\tau\to\ell$ flavour-changing interactions
  of the 125-GeV Higgs\cite{CMSLFVHiggs}:
\beq
\frac{v^2}{\LNP^2}\sqrt{{\left|{C}_{eH}^{\mu \tau}\right|}^2+{\left|{C}_{eH}^{\tau \mu}\right|}^2}
<1.00\times {10}^{-3} ~~,~~
\frac{v^2}{\LNP^2}\sqrt{{\left|{C}_{eH}^{\mathrm{e}\tau}\right|}^2+{\left|{C}_{eH}^{\tau \mathrm{e}}\right|}^2}<1.60\times {10}^{-3} ~~~.
\eeq
For $\mu\to e$ flavour change, the  MEG  bound\cite{MEG} on BR$(\meg)$  could
probe couplings as small as \cite{C+C}
\beq
\frac{v^2}{\LNP^2}{C}_{eH}^{\mu e}~~,~~~ \frac{v^2}{\LNP^2}{C}_{eH}^{e \mu} \sim  7.5 \times {10}^{-7}
\eeq
(larger values  could be  allowed if they cancel against
other contributions). These bounds imply that   in the
charged lepton mass eigenstate basis, the
 off-diagonal elements of  $Y_e$ are small
(they are comparable  to the the  LFV coefficients  $C_{eH}^{ij}v^2/\LNP^2$
---see eqn \ref{me}),    so the two largest
eigenvalues of $Y_e$ can approximately be obtained from
$m_\tau$ and $m_\mu$.
Assuming that the magnitude of the electron Yukawa
is  $\leq y_e|_{max} = m_e/v$, one obtains that in
the mass eigenstate basis, 
\bea
[Y_e] = 
 \left[ \begin{array}{ccc}
1.0\times 10^{-2} & <10^{-3} & <10^{-3} \\
 <10^{-3} & 6.0\times 10^{-4} & <10^{-6} \\
 <10^{-3} & <10^{-6} &  \leq 2.9\times 10^{-6} \\
\end{array} \right]
\approx
 \left[ \begin{array}{ccc}
\lambda^3 & \lambda^4 &  \lambda^4 \\
  \lambda^4 & 2\lambda^5 &  \lambda^9 \\
  \lambda^4 &  \lambda^9 &  \lambda^8 \\
\end{array} \right]
\label{yukawasl}
\eea

\item In  the quark  flavour sector, the mass and Yukawa matrices
select eigenbases when they are
diagonalised 
in  the generation spaces of the SM fermions.
Since this  manuscript  is focussed on  LFV,
operators such as   $H^\dagger H\bar{q} H d$   or  $H^\dagger H\bar{q} \tilde H u$
 are not considered,  and the quark masses are
 assumed to arise from Yukawa couplings. 
So the eigenvalues of $Y_d$ and $Y_u$,
evaluated at $m_W$, are taken as:
\bea
(y_b, y_s, y_d )& \approx& (1.7\times 10^{-2},3.5\times 10^{-4}, 1.7\times 10^{-5})
\approx (\lambda^2/2,\lambda^{5},\lambda^{7})
 \label{yukawas}\\
(y_t, y_c ,y_u )&\approx& (1.0,4.0\times 10^{-3}, 6.7\times 10^{-6})
\approx (1,\lambda^3/2,\lambda^7/2)
~~~. \nonumber
\eea
where  $y_f \equiv  m_f(m_W)/v$, with $m_f(m_W)$
obtained 
from one-loop RGEs --- $eg$ for quarks:
$$
m(m_W) = m(\mu)\left[\frac{\alpha_s(m_W)}{\alpha_s(\mu)}\right]^{4/\beta}
$$
with $\beta = (33-2N_f)/3 \simeq 8$, and $m(\mu)$  is from
the  PDB \cite{PDB}
with $\mu= m_b,m_c$ for
the $b,c$ and  2 GeV otherwise
\footnote{ At $m_W$, this gives
$m_b =3.0~{\rm GeV}$, $m_c =0.7~{\rm GeV}$,
$m_s =62~{\rm MeV}$, $ m_d =3.0~{\rm MeV}$,
$m_u =1.2~{\rm MeV}$.}.

The CKM matrix  is approximated in terms of $\lambda$ in usual way:
\bea
 V
 =  \left[ \begin{array}{ccc}
V_{ud} & V_{us}  &V_{ub}\\
V_{cd}  &  V_{cs}& V_{cb}\\
V_{td}  & V_{ts}& V_{tb}\\
\end{array} \right]
 =  \left[ \begin{array}{ccc} 0.974& 0.224  &-0.004\\-0.22  &  0.99\pm0.02& 0.042\\ 0.008  &-0.04 & 1.0\\ \end{array} \right]
 \simeq  \left[ \begin{array}{ccc}
1 & \lambda  &\lambda^3/2 \\
-\lambda  &  1& \lambda^2\\
\lambda^3/2  & -\lambda^2& 1\\
\end{array}
\right]
\label{CKM}
\eea

We will  always work   in the  mass eigenstate bases of
 the singlet   quarks,
 and  the $u$-type components of the doublet
 quarks. So  in the RGEs,  the up Yukawa is
 a diagonal matrix  $D_u$, and
 $Y_d = V_{CKM} D_d$. We choose the   $ \{u_L\}$ basis for quark doublets
above $m_W$   for two reasons. First,
flavour change in the RGEs is therefore
suppressed by CKM and  the small $d$-type Yukawas.
Secondly,  at dimension  six  in SMEFT,
there is only a tensor operator for $u$s (${\cal O}_{\ell equ(3)}$),
so this basis diagonalises the  large mixing of
this tensor to  the dipole operator.

The CKM matrix  is included also  in matching at $m_W$, when
the low-energy operators involving $d$-type
quarks are expressed as  SMEFT operators.

\een
The above power-counting scheme is summarised in  table
\ref{tab:pc}, and  should allow to estimate the
contribution of any operator coefficient to any
observable.

\begin{table}[ht]
\begin{center}
\begin{tabular}{|l|l|l|}
\hline
loop &$ \frac{1}{16\pi^2}$ & $\lambda^3$ \\
loop*log &$ \frac{\log }{16\pi^2}$ & $\lambda^2$ \\
lepton yukawas& $y_\tau, y_\mu ,  y_e  $ &$ \lambda^3,2\lambda^{5},\lambda^8$
\\
$\ell$ flavour change & see eqn \ref{yukawasl}&\\
$d$-quark yukawas& $y_b, y_s ,  y_d  $ &$ \lambda^2/2,\lambda^{5},\lambda^{7}$
\\
$u$-quark yukawas &$y_t, y_c ,  y_u  $ &$ 1,\lambda^3/2,\lambda^7/2$
\\
$q$ flavour change & see eqn \ref{CKM}&\\
\hline
\end{tabular}
\caption{power-counting scheme for the perturbative
expansion of the SMEFT\label{tab:pc}}
\end{center}
\end{table}

\subsection{Examples}
\label{ssec:implement}

This section gives explicit examples of how the powercounting estimates
are made, and compares them to the solutions of the RGEs.


\begin{table}[th]
\begin{center}
 \begin{tabular}{|l|l|l|}
\hline
process & bound  on BR  & sensitivity to $C$    \\
\hline
$\meg $
& $ < 4.2 \times 10^{-13}$ \cite{MEG} 
$\to  6 \times 10^{-14}$ \cite{MEGII}  &  $C_D\sim
\lambda^{11} \to \lambda^{12}$\\
%
$\meee $
& $ < 1.0  \times 10^{-12}$ \cite{Bellgardt:1987du}
$\to  10^{-16}$ \cite{Mu3e}
&
$C_S\sim \lambda^{8} \to \lambda^{11}$ \\
 && $C_V\sim \lambda^{8.5} \to \lambda^{11.5}$ \\
$\mu A \to e A$
&  $< 7 \times 10^{-13}$ \cite{Bertl:2006up} 
$ \to   10^{-16}$ \cite{Mu2e,COMET}
   &   $C_{V,D} \sim \lambda^{9.5} \to \lambda^{12}$ \\
&     & $C_{S} \sim \lambda^{10.5} \to {\lambda^{14}}$ \\
\hline
 $\overline{K^0_L} \to \mu \bar{e}$ &$ < 4.7 \times 10^{-12}$   &
 $C_P  \sim \lambda^{11.5}$ \\
 && $C_A  \sim \lambda^{9.5}$\\
\hline
 $B^0_d \to \mu^\pm e^\mp$ &$ < 1 \times 10^{-9}$   &$C_P  \sim \lambda^{7.5}$\\
$B^+_d \to \pi^+ \bar{\mu} e$ &$ <1.7  \times 10^{-7}$ 
&$C_V  \sim \lambda^{7}$ \\ \hline
  $B^0_s \to \mu^\pm e^\mp$ &$ < 5.4 \times 10^{-9}$   &$C_P  \sim \lambda^{7.5}$\\
$B^+ \to K^+ \bar{\mu} e$ &$< 9.1 \times 10^{-8}$ 
& $C_V  \sim \lambda^{6.5}$\\
\hline
  $D^0 \to \mu^\pm e^\mp$ &$ < 1.3 \times 10^{-8}$   &$C_P  \sim \lambda^{6}$\\
$D^+ \to \pi^+ \bar{\mu} e$ &$<1.7  \times 10^{-7}$ 
&$C_V  \sim \lambda^{4}$\\
\hline
\hline
$\tau\to \ell\gamma$ & $<3.3\times 10^{-8}$ \cite{tau1} & $C_D\sim \lambda^{7.5}$\\
$\tau\to \ell\bar{\ell} \ell$ & $\lsim 2\times 10^{-8}$ \cite{tau2}
$\to \lsim 10^{-9} $\cite{belle2t3l}
   & $C_V\sim \lambda^{5}\to \lambda^{5.5}$ \\
   & &  $C_S\sim \lambda^{4.5}\to \lambda^{5}$\\
\hline 
$\tau\to \ell\pi^0$ & $<8.0\times 10^{-8}$ \cite{tau3} & $C_S\sim \lambda^{4.5}$\\
$\tau\to \ell\eta$ & $<6.5\times 10^{-8}$ \cite{tau3} & $C_S\sim \lambda^{4.5}$\\
$\tau\to \ell\rho$ & $<1.2\times 10^{-8}$ \cite{tau4} & $C_V\sim \lambda^{4.5}$\\
\hline
             $B^0_d\to e\tau$ & $<2.8\times 10^{-5}$ \cite{PDB}& $C_{P}\sim \lambda^5$\\
              & & $C_{A}\sim \lambda^{4.5}$\\
              \hline
\end{tabular}
\end{center}
\caption{ Some current and upcoming experimental bounds on
LFV  Branching Ratios ($\tau \leftrightarrow \mu$ results
are similar to $\tau \leftrightarrow e$).
The third colomn gives the order of  magnitude of
 dimension six operator coefficients that
 reproduce the experimental numbers, 
in  powers of  $\lambda\simeq 1/5$. 
The listed  coefficients 
$C_{Lor}$   contribute to the process
at tree level, are labelled by
the operator's Lorentz structure, and
 are normalised to a  scale
 $\Lambda  = v= 174$  GeV  in eqn (\ref{L}). The meson decay
bounds are from \cite{PDB}, the coefficient
sensitivities from \cite{AS,CD,C+C}.
\label{tab:exptme} }
 \end{table}

We first consider $\mu \to e$ processes because
the most restrictive experimental  constraints on LFV
arise in this sector, and
upcoming experiments aim to improve the
sensitivities by several orders of magnitude
(see table \ref{tab:exptme}; indeed, there
plans to reach a conversion ratio  $\lsim 10^{-18}$
for $\muc$ \cite{PP}).
The Branching Ratios 
can be expressed  (see {\it eg}\cite{KO,C+C,AS})
in terms of  the  coefficients,
evaluated at the experimental scale,
of operators which contribute at tree level.
For instance, the low-energy operators
\beq
\delta {\cal L} = 2\sqrt{2} G_F (C_{D,L} m_\mu \overline{e}\sigma \cdot F P_L \mu +   C_{D,R} m_\mu \overline{e}\sigma \cdot F P_R \mu)
\label{D1}
\eeq
 contribute to $\meg$ \cite{MEG} as
 \beq
BR(\meg) = 384\pi^2 (|C_{D,R}|^2 + |C_{D,L}|^2) < 4.2\times 10^{-13}
\eeq
which  gives the experimental bounds, translated into
our power counting parameter ($\Lambda \sim v$ in eqn (\ref{L})
\beq
 |C_{D,R}|, |C_{D,L}| < 1.05\times 10^{-8} \sim \lambda^{11} ~~~.
\eeq
The dipole is a special case, because the operators
contain not only fields, but also a built-in parametric
suppression factor $m_\mu$. This is the usual operator definition,
and makes sense because  in SMEFT the  operator has a Higgs
leg which  frequently  attaches to the muon line.
However, in some loop diagrams (for instance Barr-Zee)
the Higgs  is attached to a heavier particle in a loop,
so such diagrams would  gain  a factor $1/(2\lambda^5)$ in our power-counting scheme. 
For a different normalisation of the dipole operator, the power-counting
sensitivity would change. For instance,
\beq
\delta {\cal L} = 2\sqrt{2} G_F (C_{D,L} v \overline{e}\sigma \cdot F P_L \mu +   C_{D,R} v \overline{e}\sigma \cdot F P_R \mu)
\label{D2}
\eeq
gives  $ |C_{D,R}|, |C_{D,L}| \lsim \lambda^{16} $.

\begin{figure}[ht]
\unitlength.50mm
\SetScale{1.4}
%
\hspace{2cm}
\begin{picture}(80,60)(-40,-10)
\DashLine(-30,50)(-14,34){3}
\ArrowArc(0,20)(20,0,360)
\ArrowLine(20,20)(40,40)
\ArrowLine(40,0)(20,20)
\Photon(-40,20)(-20,20){2}{4}
\GCirc(20,20){5}{.7}
\Text(20,20)[c]{T}
\Text(48,46)[c]{$e$}
\Text(48,-6)[c]{$\mu$}
\Text(-5,46)[c]{$f$}
\Text(-5,-6)[c]{$f$} 
\end{picture}
\begin{picture}(80,60)(-30,-10)
\ArrowLine(0,0)(20,20)
\ArrowLine(20,20)(0,40)
\ArrowLine(20,20)(40,40)
\ArrowLine(40,0)(20,20)
\Photon(35,5)(5,5){2}{4}
\GCirc(20,20){5}{.7}
\Text(20,20)[c]{T}
\Text(48,46)[c]{$e$}
\Text(48,-6)[c]{$\mu$}
\Text(-5,46)[c]{$f$}
\Text(-5,-6)[c]{$f$}
\Text(35,20)[l]{ $+ ...~~~\Rightarrow$}
\end{picture}
\hspace{1cm}
\begin{picture}(80,60)(-10,-10)
\ArrowLine(0,0)(20,20)
\ArrowLine(20,20)(0,40)
\ArrowLine(20,20)(40,40)
\ArrowLine(40,0)(20,20)
\GCirc(20,20){5}{.7}
\Text(20,20)[c]{S}
\Text(48,46)[c]{$e$}
\Text(48,-6)[c]{$\mu$}
\Text(-5,46)[c]{$f$}
\Text(-5,-6)[c]{$f$}
\end{picture}
\caption{On the left, a diagram mixing the tensor operator to the dipole (the Higgs leg is replaced by a mass insertion in the EFT below $m_W$).
On the right, one of the diagrams mixing tensor operators to scalars (the gauge boson can attach to any two legs not belonging to the same bilinear).
\label{fig:T->D}}
\end{figure}
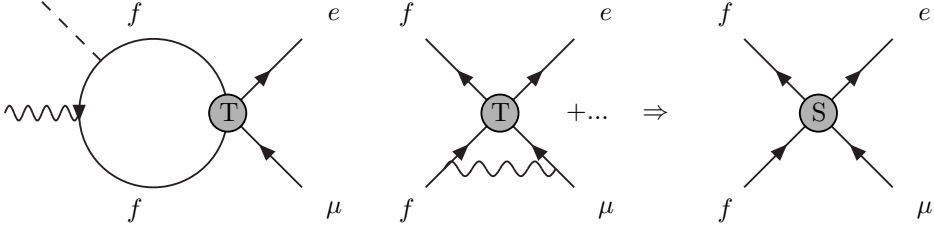

\begin{figure}[ht]
\unitlength.50mm
\SetScale{1.4}
%
\hspace{2cm}
\begin{picture}(120,70)(0,-10)
				\ArrowLine(0,0)(20,0)
				\ArrowLine(20,0)(70,0)
				\ArrowLine(70,0)(90,0)
				\DashLine(35,0)(45,-20){2}
				\ArrowArc(60,20)(17,-120,240)
				\Photon(60,37)(60,60){2}{4}
				\Photon(43,18)(20,0){2}{4}
				\GCirc(60,0){7}{.9}
				\Text(60,0)[c]{V}
				\Text(5,-5)[r]{$\mu_R$}
				\Text(82,-5)[l]{$e_L$}
				\Text(28,-10)[c]{$y_\mu$}
				\Text(49,60)[l]{$\gamma$}
				\Text(40,37)[l]{$q_L$}
			\end{picture}
	\begin{picture}(120,70)(0,0)
				\ArrowLine(0,0)(20,0)
				\ArrowLine(20,0)(70,0)
				\ArrowLine(70,0)(90,0)
				\DashLine(85,45)(75,30){2}
				\ArrowArcn(60,17)(17,0,180)
                                \ArrowArcn(60,17)(17,180,270)
				\Photon(60,35)(60,60){2}{4}
				\Photon(43,18)(20,0){2}{4}
				\GCirc(60,0){7}{.9}
				\Text(60,0)[c]{V}
				\Text(5,-5)[r]{$\mu_R$}
				\Text(82,-5)[l]{$e_L$}
				\Text(64,25)[l]{$y_d$}
				\Text(47,60)[l]{$\gamma$}
				\Text(78,15)[l]{$q_L$}
				\Text(42,35)[l]{$d_R$}
			\end{picture}
\caption{Representative diagrams allowing
two-loop   mixing  of vector operators 
		to the dipole. 
\label{fig:V->D}}
\end{figure}
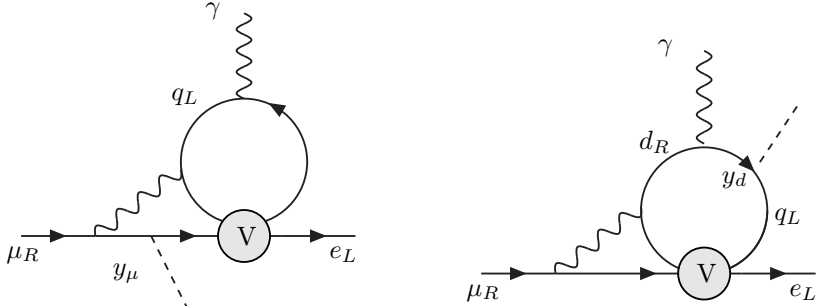

The sensitivity of $\meg$ to  other operators can be estimated in our power-counting  scheme  by drawing diagrams.  For instance,
tensor operators  mix to  the dipole via the
left  diagram  of Fig. \ref{fig:T->D}.
Below the electroweak scale and normalizing as in eq.~(\ref{D1}),
the contribution to the dipole coefficient is of order
\begin{equation}
	\Delta C_D\frac{m_\mu}{v^2}\sim e\frac{\log}{16\pi^2} C^{ff}_T\frac{m_f}{\Lambda_{\rm NP}^2} \Rightarrow \Delta C_D\sim e\lambda^2 C^{ff}_T \frac{y_f}{y_\mu}\frac{v^2}{\Lambda_{\rm NP}^2}
        \label{DDdeT}
\end{equation}
where $f=u,d,s,c,b,e,\mu,\tau$, and
the estimate in our power-counting scheme can be obtained
using table \ref{tab:pc}.
\begin{table}[h]
\begin{center}
\begin{tabular}{|ccc|}
	$f$ & Power Counting & Running\\
	\hline
	$ e $& $\sim 20\ \rm{T}e\rm{V}$ & $\sim13\ \rm{T}e\rm{V}$ \\
	$ \mu $& $\sim 300\ \rm{T}e\rm{V}$ & $\sim190\ \rm{T}e\rm{V}$ \\
	$ \tau $& $\sim 10^3\ \rm{T}e\rm{V}$ & $\sim1.1\times 10^3\ \rm{T}e\rm{V}$ \\
	$ u $& $\sim 50\ \rm{T}e\rm{V}$ & $\sim71\ \rm{T}e\rm{V}$ \\
	$ d $& $\sim 50\ \rm{T}e\rm{V}$ & $\sim73\ \rm{T}e\rm{V}$ \\
	$ s $& $\sim 200\ \rm{T}e\rm{V}$ & $\sim330\ \rm{T}e\rm{V}$ \\
	$ c $& $\sim 10^3\ \rm{T}e\rm{V}$ & $\sim1.7\times 10^3\ \rm{T}e\rm{V}$ \\
	$ b $& $\sim 2\times10^3\ \rm{T}e\rm{V}$ & $\sim2\times 10^3\ \rm{T}e\rm{V}$ \\
\end{tabular}
\end{center} 
\caption{Powercounting estimates of the mixing from tensor to dipole operators below $m_W$, compared to the solutions of   the RGEs \cite{C+C,PSI}.\label{tab:TtoD}}
\end{table}


Scalar and vector operators can contribute to the dipole  via two-loop
diagrams, that arise either as one-loop mixing into the tensor, or
direct mixing   to the dipole at two-loop.  
Below the weak scale,  the scalar  to tensor  mixing is via  diagrams like 
the right  figure \ref{fig:T->D}, where the gauge boson is a photon,  which gives
\begin{equation}
	\Delta C_D \sim e^3 \frac{\log^2}{(16\pi^2)^2}C_S^{ff}\frac{y_f}{y_\mu} \frac{v^2}{\Lambda_{\rm NP}^2}
        \label{S->D}
\end{equation}
where now $f=u,d,s,c,b,\tau$. The vector to dipole mixing is via diagrams such as figure  \ref{fig:V->D}. We estimate  the diagrams on the left and
right as
\begin{equation}
	\Delta C_D \sim e^3 \frac{\log}{(16\pi^2)^2}C_V\left(\frac{v}{\Lambda_{\rm NP}}\right)^2 \times \left\{
        \begin{array}{c}
        1\\
        \frac{y_d}{y_\mu}
        \end{array}
        \right.
        \label{V->D}
\end{equation}
so there is sensitivity to vector coefficients for scales
below 10 T$e$V (which is consistent with the bound in \cite{C+C,PSI}).

Comparing  to the  solution
of 1-loop  RGEs below the electroweak scale \cite{C+C,PSI},
we  find  that the scale $\LNP$ where $C_T^{ff}$ would be $\sim 1$,
obtained by combining eqn (\ref{DDdeT}) with column three
of table \ref{tab:exptme}, 
differs by at most $\sqrt{3}$   from the solution of the  RGEs,
as can be seen in table \ref{tab:TtoD}.
For the  second-order/two-loop mixing of eqns  (\ref{S->D},\ref{V->D})
we find  that  the powercounting   can mis-estimate $\LNP$ 
by a factor 2-3 (which corresponds to a factor 4-9 in $C$).

\section{Questions}
\label{sec:qestions}

This section  uses the power-counting proposal of the previous section
to study what physics should be
included at ``leading order'', in  the  SMEFT RGEs for 
LFV operators. In the first sections, the focus is
on $\mu \leftrightarrow e$ flavour change,
due to the sensitivity of current and upcoming experiments;
the importance of dimension eight operators
and  two-loop  anomalous dimensions for $\tau$-LFV  is briefly discussed in
section \ref{ssec:tau}.

\subsection{Dimension eight operators}
\label{ssec:dim8}

This section explores 
when which dimension eight  operators  are required,
and whether their  RGEs are  required.

We suppose  that the New Physics responsable for
LFV is beyond the reach of the LHC, so $\LNP\gsim$ 4 TeV. 
In the normalisation convention of   table  \ref{tab:exptme},
this implies that  coefficients of dimension eight operators  at 
 are suppressed by $\sim \lambda^8$:
\beq
\frac{f(g_{NP},...)}{(4~{\rm TeV})^4} {\cal O}^{(8)} = \frac{^{(8)}C}{v^4} {\cal O}^{(8)}
\Rightarrow ~^{(8)}C \lsim \frac{v^4}{(4~{\rm TeV})^4} \simeq \lambda^8
\label{eqnn}
\eeq
Comparing to the tree-level sensitivities  given in table
\ref{tab:exptme}, one sees  that kaon and muon  decays
are generically sensitive to dimension eight operators
induced by new particles in the
interesting mass range just beyond the reach of the LHC.
Pushing the  New Physics scale
above 20 TeV would give $^{(8)}C\lsim \lambda^{12}$,
making most dimension eight operators irrelevant.

There are  thousands of  LFV dimension eight operators
\cite{LRSXYZ,Murphy:2020rsh},
so  it would be  attractive to include only some of them
in a first approximation. Indeed, in a
bottom-up perspective, only the dimension eight operators to which operators
are sensitive are required.
So we reject derivative operators such as  
$$
D^\a(\overline{e}\g_\b \mu )D_\a(\overline{f}\g^\b f )
$$
because their contribution  to low-energy S-matrix elements
should be suppressed by  $\{s,t,u\}/v^2$,
suggesting that $K$ and $\mu$ processes  have no
sensitivity to dimension eight derivative operators.
We also neglect operators with more than four legs after
electroweak symmetry-breaking,
on the assumption that they  do not contribute
(at tree level) to our low-energy observables.

There remain about four dozen  $\mu \leftrightarrow e$
operators (given
 in  Appendix \ref{app:O8} in the notation of \cite{Murphy:2020rsh}):
\ben

\item   four-particle   operators  which are forbidden at
  dimension 6 due to gauge invariance.

\item dimension six SMEFT  operators   with an
additional  $H$ and $H^\dagger$, such
as $(HH^\dagger)\bar\ell H\sigma^{\alpha\beta}eF_{\alpha\beta}$
or $ (\bar\ell_eH\sigma^{\alpha\beta} \mu)(\bar q_i\tilde{H}\sigma_{\alpha\beta}u_j)
$. It may seem unlikely that
the dimension eight contribution could be relevant
given the possibility of a dimension six term\footnote{
The dimension six coefficient
  could perhaps be  suppressed by additional loops or small
  couplings with respect to dimension eight.};
however,  being agnostic could be appropriate in EFT, and
dimension eight operators are considered, for instance,
in studies of Non-Standard neutrino Interactions
\cite{NSI}.

\een
These operators  are schematically listed in tables
\ref{tab:d8LNP>} and \ref{tab:d8LNP>2},  along with
the scale below  they could contribute to
observables with a coefficient $C\lsim 1$.
So they should be considered  in the
 EFT parametrisation of  any model 
constructed below this scale.

\begin{table}[ht]
\begin{center}
 \begin{tabular}{|l|c|l|}
\hline
operator &  $\LNP$ (in TeV)   & process     \\
\hline
$ (\overline{\ell}_eH e_\mu)(\overline{q}_d {H} d_d)$
&55 & $\muc$\\
$ (\overline{\ell}_eH e_\mu)(\overline{u}_u \tilde{H}^\dagger q_u)$
&55 & $\muc$\\
$ (\overline{\ell}_eH e_\mu)(\overline{q}_s {H} d_s)$
&26 & $\muc$\\
$ (\overline{\ell}_eH \sigma  e_\mu)(\overline{q}_d {H}\sigma  d_d)$
&25 & $\muc$\\
$ (\overline{\ell}_eH \sigma  e_\mu)(\overline{q}_b {H}\sigma  d_b)$
&25 & $\muc$\\
$ (\overline{\ell}_eH e_\mu)GG$
&20 & $\muc$\\
$ (\overline{\ell}_eH\sigma e_\mu)(\overline{\ell}_\tau{H} \sigma e_\tau)$
&20 & $\meg$\\
$ (\overline{\ell}_eH e_\mu)(\overline{\ell}_e {H} e_e)$
&15 & $\meee$\\
$ (\overline{\ell}_eH e_\mu)(\overline{u}_c \tilde{H}^\dagger q_c)$
&15 & $\muc$\\
$ (\overline{\ell}_eH \sigma  e_\mu)(\overline{q}_s {H}\sigma  d_s)$
&15 & $\muc$\\
$ (\overline{\ell}_eH  e_\mu)(\overline{u}_t\tilde{H}^\dagger   q_t)$
&10 & $\meg$\\
$ (\overline{\ell}_eH   e_\mu)(\overline{q}_b {H} d_b)$
&10 & $\muc$\\
$ (\overline{\ell}_eH e_\mu)(\overline{\ell}_\mu {H}  e_\mu)$
&8 & $\meg$\\
$ (\overline{\ell}_eH e_\mu)FF$
&3 & $\muc$\\
\hline
\end{tabular}
\end{center}
\caption{ Dimension eight operators which induce at low energy
  four-particle contact interactions  that do not arise at dimension six.
  The operators are represented schematically in the first colomn,
  and   the second colomn gives the scale  $\LNP$ up to which the  process
  of the third colomn (with upcoming sensitivity)  could probe coefficients $\lsim 1$. (The estimate for   $ (\overline{\ell}_eH e_\mu)FF$ is from \cite{DKUY}.)
\label{tab:d8LNP>}.}
 \end{table}

The effects of these operators can be partially accounted
for  by  matching the model onto them at $\LNP$,
and then including them in
the matching at the weak scale onto the low energy EFT.  These 
matching conditions for LFV operators
are given in appendix \ref{app:match} (at tree level).

Many of these operators contribute to observables via loops,
so including them in RGEs is relevant.
Since they match at $m_W$ onto low-energy four-particle
interactions, the
Renormalisation Group running below $m_W$ is known 
and will occur automatically once they are included
in the matching.

The RG running in SMEFT is missing.
 Above $m_W$ , the Higgs and $W$
bosons can   mix operators differently from
the gluon and photon, for instance by modifying the SU(2) contractions
(see eg  the RGEs for a subset of dimension eight
operators in \cite{NSI3}).
 Dimension eight  four-fermion operators involving two  tops pose
 a particular problem, because their leading contribution to
 low energy LFV is likely to arise from  the unknown RG running in SMEFT.
 Fortunately, many of these top operators are
dimension six operators with  an extra 
$H^\dagger$ and  $H$ (only  the  operator
$\sim (\overline{e} P_R \mu)(\overline{t}P_L t)$ 
arises first at dimension eight), so one could
hope that models dominantly  generate  dimension six operators.
Alternatively, one could  envisage to
add the  coefficients of dimension eight top
operators to the dimension six coefficients
at $\LNP$, and evolve them with the
 SMEFT  RGEs at dimenson six, which  will
 include a subset of the loops.
 We leave calculating the anomalous dimensions for
 a later project.

\begin{table}[ht]
\begin{center}
 \begin{tabular}{|l|c|l|}
\hline
operator &  $\LNP$ (in TeV)   & process     \\
\hline
$ (H^\dagger H)  (\overline{\ell}_e\sigma e_\mu)(\overline{q}_t \sigma u_t)$& 100&$\meg$ \\
$(HH^\dagger)(\bar\ell_e e_\mu) (\bar d_d q_d)$& 55 & $\muc$\\
$(HH^\dagger)(\bar\ell_e e_\mu)\epsilon (\bar q_u u_u)$& 55 & $\muc$\\
$(HH^\dagger)(\bar\ell_e e_\mu) (\bar d_s q_s)$& 25 & $\muc$\\
$(HH^\dagger)(\bar\ell_e \gamma^{\alpha}\ell_\mu) (\bar q_u \gamma_{\alpha}q_u)$& 22 & $\muc$\\
$(HH^\dagger)(\bar\ell_e \gamma^{\alpha}\ell_\mu) (\bar u_u \gamma_{\alpha}u_u)$& 22 & $\muc$\\
$(HH^\dagger)(\bar\ell_e \gamma^{\alpha}\ell_\mu) (\bar q_d \gamma_{\alpha}q_d)$& 22 & $\muc$\\
$(HH^\dagger)(\bar\ell_e \gamma^{\alpha}\ell_\mu) (\bar d_d \gamma_{\alpha}d_d)$& 22 & $\muc$\\
$(HH^\dagger)\bar\ell_e H\sigma^{\alpha\beta}e_\mu F_{\alpha\beta}$& 20 & $\meg$\\
$(HH^\dagger)(\bar\ell_e \gamma^{\alpha}\ell_\mu) (\bar\ell_e \gamma_{\alpha}\ell_e)$& 18 & $\meee$\\
$(HH^\dagger)(\bar\ell_e \gamma^{\alpha}\ell_\mu) (\bar e_e \gamma_{\alpha}e_e)$& 18 & $\meee$\\
$(HH^\dagger)(\bar e_e \gamma^{\alpha}e_\mu) (\bar e_e \gamma_{\alpha}e_e)$& 18 & $\meee$\\
$(HH^\dagger)(\bar\ell_e e_\mu)\epsilon (\bar q_c u_c)$& 15 & $\muc$\\
$(HH^\dagger)(\bar\ell_e e_\mu) (\bar d_b q_b)$& 10 & $\muc$\\
\hline
\end{tabular}
\end{center}
\caption{  Dimension eight operators which induce  low energy
  contact interactions  that {\it do } arise at dimension six.
  In the first colomn the operators are represented schematically(other
  distributions of the Higgses, or triplet constractions, could be possible),
  and   the second colomn gives the scale  $\LNP$ up to which the  process
  of the third colomn (with upcoming sensitivity)  could probe coefficients $\lsim 1$. 
  \label{tab:d8LNP>2}.}
 \end{table}

\subsection{2-loop anomalous dimensions?}
\label{ssec:2loop}

This section  aims to identify  relevant mixing that
could arise from the two-loop RGEs of SMEFT, so we
are looking for two-loop  diagrams that  would {\it not}
be generated at second order in the one-loop RGEs.

One can  see why these could be interesting, 
by considering the QED$\times$QCD-invariant EFT below $m_W$,
where   at one-loop, vector operators mix among themselves, and
the dipoles+scalars+tensors mix among themselves,
but there are no divergent one-loop diagrams 
mixing  vectors and  non-vectors.
Therefore, to all orders in the one-loop RGEs,
the vectors evolve  separately from the others. 
However, vector
to dipole mixing occurs at
two-loop,  and is encoded in the the two-loop RGEs \cite{luca};
a few  diagrams are given in figure \ref{fig:V->D}.
So we are looking for two-loop diagrams that
allow operator  ${\cal O}$ to mediate process
${\cal P}$,  when   ${\cal O}$ cannot mediate
${\cal P}$ via  the one-loop RGEs.

\begin{figure}[ht]
	\unitlength.5mm
	\SetScale{1.4}
	%
	\hspace{4cm}
	\begin{picture}(45,50)(0,0)
		\ArrowLine(0,0)(20,20)
		\ArrowLine(20,20)(0,40)
		\ArrowLine(20,20)(40,40)
		\ArrowLine(40,0)(20,20)
		\DashArrowLine(5,5)(35,5){2}
		\Text(20,20)[c]{V}
		\GCirc(20,20){5}{.9}
		\Text(20,20)[c]{V}
		\Text(48,46)[c]{$e_L$}
		\Text(48,-6)[c]{$\mu_R$}
		\Text(-5,46)[c]{$q_L$}
		\Text(-5,-6)[c]{$u_R$}
		\Text(50,20)[l]{ $~~~\Rightarrow$}
	\end{picture}
	\hspace{3cm}
	\begin{picture}(45,40)(0,0)
		\ArrowLine(0,0)(20,20)
		\ArrowLine(20,20)(0,40)
		\ArrowLine(20,20)(40,40)
		\ArrowLine(40,0)(20,20)
		\GCirc(20,20){5}{.9}
		\Text(20,20)[c]{T}
		\Text(48,46)[c]{$e_L$}
		\Text(48,-6)[c]{$\mu_R$}
		\Text(-5,46)[c]{$q_L$}
		\Text(-5,-6)[c]{$u_R$}
	\end{picture}
	\caption{Vector mixing to the  tensor via Higgs exchange.\label{fig:V>T}}
\end{figure}
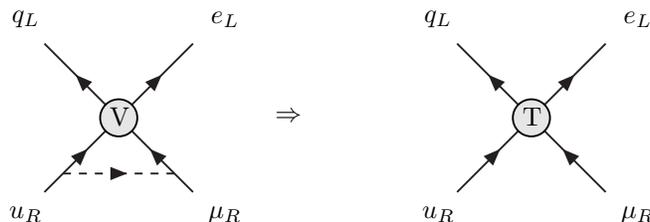

In SMEFT, there can be 1-loop vector to tensor mixing  by exchanging an Higgs, as illustrated in Fig.\ref{fig:V>T}. Closing the quark legs  gives  a contribution to the dipole. For instance, considering the vector $\mathcal{O}^{(1)}_{\ell q}$ we find

\begin{equation}
\Delta C_D\sim e   \left(\frac{\log}{16\pi^2}\right)^2 C_{\ell q}^{(1) e \mu nm} [Y_u Y_u^\dagger]_{nm} \frac{v^2}{\LNP^2}
\sim e \lambda^4  C_{\ell q}^{(1) e \mu nm} [Y_u Y_u^\dagger]_{nm} \frac{v^2}{\LNP^2}
\label{num}
\end{equation}
which results in a sensitivity to $ C_{\ell q}^{(1) e \mu tt}$ up to $\Lambda_{\rm NP}\sim 50$ T$e$V. Estimates similar to eqn (\ref{num})
hold for all vector operators  which can mix to the $u$-type tensor.

Vector operators also  can mix directly  to the dipole in the 2-loop RGEs through gauge interactions, as illustrated by the diagram  on the left of figure  \ref{fig:V->D}.
{ The powercounting estimate for these diagrams
\begin{equation}
	\Delta C_D \sim e^3 \frac{\log}{(16\pi^2)^2}C_V\frac{v^2}{\Lambda^2_{\rm NP}}
\end{equation}
suggests that there is sensitivity to vector coefficients for scales
below 10 T$e$V 
---which  is larger than  the vector$\to$tensor$\to$ dipole
contribution for all operators not involving a top quark,
see table \ref{tab:2l}.
}

There could also  be two-loop mixing of  the    ${\cal O}_{LEDQ}$ scalar to
the dipoles.  For comparaison, at one loop
the  $u$ quark scalar operator
${\cal O}_{LEQU}$ mixes  to the tensor,
which mixes to the dipole, and  due to Yukawa
enhancement and large anomalous dimensions,
this  second-order process in the one-loop
RGEs is important.  In the $d$-quark sector, there
is no dimension six tensor, so no equivalent process
occurs; however the diagrams are there,
and ${\cal O}_{LEDQ}$  can be Fierzed
to the vector $-\frac{1}{2}(\overline{\ell} \g_\a q) (\overline{d} \g_\a e)$
which mixes  at two-loop to the dipole \cite{luca}.
The powercounting  estimate is
\begin{equation}
	\Delta C_D\sim eg^2 \lambda^5 C^{ij}_{\ell edq}[Y_d]_{ij}\frac{v^2}{\Lambda_{\rm NP}^2}
\end{equation}
which suggests that  $\meg$  could
be sensitive  to coefficients $\lsim 1$ up to the scales
given in table \ref{tab:2l}.

\begin{table}[ht]
\begin{center}
 \begin{tabular}{|l|c|c|}
\hline
operator & 2loop  V$\to$ D($\LNP$ in TeV)   &  V$\to$T$\to$D($\LNP$ in TeV)      \\
\hline
${\cal O}_{\ell q}^{(1) e\mu tt}$&10 &50\\
${\cal O}_{V}^{ e\mu ff}$&10 &\\
${\cal O}^{e\mu dd}_{\ell e d q}$&5&---\\
${\cal O}^{e\mu ss}_{\ell e d q}$&20  &---\\
${\cal O}^{e\mu bb}_{\ell e d q}$&100&---\\
\hline
\end{tabular}
\end{center}
\caption{Operators which contribute to $\meg$ via two-loop mixing in the RGEs,
and  in the second colomn, our powercounting estimate for 
the scale $\LNP$  up to which coefficients $\lsim1$ could be probed. 
The third colomn gives the  estimated sensitivity  obtainable via
 the one-loop RGEs.  ${\cal O}_{V}^{ e\mu ff}$ schematically refers to
all the dimension six vector four-fermion operators with
$f \neq t$.
\label{tab:2l}.}
 \end{table}

These results show  that the two-loop vector to dipole mixing can be relevant,
and often dominates over the  mixing involving  a Higgs loop, 
which occurs at second-order in the  one-loop  RGEs.  It would be
desirable to include these two-loop anomalous dimensions.
However, although they are known in QCD and QED \cite{luca,czar,PSI},
a complete computation in SMEFT is currently missing in the literature
\cite{SJL}.

\subsection{CKM}
\label{ssec:CKM}

  CKM mixing angles can appear in various  places
in SMEFT:  in matching of the higher scale theory onto SMEFT,
in the RG running of  operator coefficients and of  SM
couplings,  and in matching the SMEFT operators at
$m_W$ onto the  QED$\times$QCD-invariant  low energy
 theory. Including CKM in matching  at $m_W$ is straightforward, 
but it could be conceptually simpler
to  set $V_{CKM} = 1$ in the RGEs
for the operator coefficients.
This section explores the  errors
that could arise from this approximation,
by allowing  one
non-zero operator at a time at $\LNP$, and estimating the magnitude
of low-energy  coefficients that   it generates   at one-loop
$\propto  [V_{CKM}]_{ij}, i\neq j$. If no experiment has
sensitivity to  the contributions  proportional
to CKM mixing angles, then one can conclude that
$V_{CKM}=1$ is an acceptable approximation
in the RGEs.

 The CKM matrix also appears in the RGEs of the renormalisable SM couplings,
where it causes the eigenbases of $Y_d Y_d^\dagger$
and $Y_uY_u^\dagger$ to rotate with scale.
This is due to wavefunction corrections.
Since  wavefunction diagrams also decorate the operators,
we  assume this is a ``universal'' effect,
automatically  included by working in  the
rotating  $Y_uY_u^\dagger$ eigenbasis, and do not powercount
 the associated diagrams \footnote{ For
 instance, an off-diagonal
 $[Y_u]^{ct}\sim 3\log/(32\pi^2) V_{cb}y_b^2V_{tb}y_t$
 is generated by  a Higgs  loop on the $q_L$ line.
 Inside the  loop mixing ${\cal O}_{LEQU,3}\to {\cal O}_{D}$,
 this could give  sensitivity to  ${\cal O}_{LEQU,3}^{e\mu ct}$,
 in an unrotating basis for $q_L$.}.

Recall that we work in the $Y_d$ eigenbasis for the $\{ d_R\}$,
and the  $Y_u$ eigenbasis for the $\{ u_R\}$ and  $\{ q_L\}$. So
$V_{CKM}$ only appears  in Higgs loops, at vertices $\propto  Y_d = V_{CKM} D_d$.
It therefore enters the one-loop  RGEs of
${\cal O}_{LQ1}$,${\cal O}_{LQ3}$, ${\cal O}_{LD}$, ${\cal O}_{ED}$
${\cal O}_{EQ}$ and ${\cal O}_{LEDQ}$.

Consider first   operators  at $\LNP$ with a doublet quark bilinear
$(\overline{q}_i \g_\a q_j)$, where $i,j \in\{u,c,t\}$.
Higgs exchange between the quark legs can dress
this quark bilinear  to generate 
\bea
(\overline{q}_i \g_\a q_j)
 \to  V^{ip} y_{d_p} V^{jr}y_{d_r} \frac{\log}{16\pi^2 } (\overline{d}_p \g_\a d_r)
\eea
where  the  approximate magnitude of $V^{ip} y_{d_p} V^{jr}y_{d_r}$, for all  possible flavours of the  doublet and singlet lines,
is given in table \ref{tab:qq->dd}.  If the  CKM matrix  is approximated as the identity, then only the diagonal components of the table would remain.

\begin{table}[ht]
\begin{center}
\begin{tabular}{|l|c|c|c|c|c|}
\hline
$ij\backslash pr$& bb&bs&ss&sd&dd\\
\hline
tt&$\lambda^{5}$&$\lambda^{9.5}$&$\lambda^{14}$&$\lambda^{17.5}$&$\lambda^{21}$\\
tc&$\lambda^{7}$&$\lambda^{7.5}$&$\lambda^{12}$&$\lambda^{15}$&$\lambda^{18.5}$\\
cc&$\lambda^{9}$&$\lambda^{9.5}$&$\lambda^{10}$&$\lambda^{13}$&$\lambda^{16}$\\
cu&$\lambda^{10.5}$&$\lambda^{10.5}$&$\lambda^{11}$&$\lambda^{12}$&$\lambda^{15}$\\
uu&$\lambda^{12}$&$\lambda^{12}$&$\lambda^{12}$&$\lambda^{13}$&$\lambda^{14}$\\
\hline
\end{tabular}
\caption{Estimates for the  Yukawa and CKM suppression
($\simeq  V^{ip} y_{d_p} V^{jr}y_{d_r}$) of the mixing
between  operators
containing $(\overline{q}_i \g_\a q_j)$ into operators
containing $(\overline{d}_p \g_\a d_r)$. The indices
$ij$ are given in the left colomn, and $pr$ in the
top line. 
\label{tab:qq->dd}}
\end{center}
\end{table}

From the  table \ref{tab:qq->dd}, one sees
that  mixing induced by  non-vanishing CKM angles
is suppressed by    $ <\lambda^{7+2} v^2/\LNP^2$
(where the additional $\lambda^2$ is  for
the $\log/16\pi^2$ loop suppression).
Such contributions  are clearly negligeable
in the RGEs for  $\tau\to \ell$
 operators; 
to determine whether
 they should be included in
 the RGEs for  $\mu \leftrightarrow e$  operators,
 we compare to the sensitivity of
 upcoming experiments.
In the case of $p=r$ but $i\neq j$,
the best sensitivity  is from $\mec$. 
We  estimate  that $\muc$  could be
sensitive to the mixing  from    $(\overline{q}_t \g_\a q_c) \to
 (\overline{b} \g_\a  P_R b)$
 for an experimental reach  BR($\muc) \lsim 10^{-16}\frac{v^4}{\LNP^4}$,
 and to   the $cu\to ss,dd$ mixing  for BR($\muc) \lsim 10^{-20}\frac{v^4}{\LNP^4}$. 
This suggests that the RGE-mixing of 
operators involving  $(\overline{q}_i \g_\a q_j)$, 
into 
operators involving
 $(\overline{d}_p \g_\a d_p)$, for $i\neq j $   and
 $p=q$,  is negligeable in the forseeable future. 
In the converse case, of  RGE-mixing of  flavour-diagonal
operators   $(\overline{q}_i \g_\a q_i)$, 
into  quark flavour non-diagonal operators
$(\overline{d}_p \g_\a d_r)$,   table 
\ref{tab:qq->dd} indicates that the least suppressed
mixings  are  $tt,cc\to bs\propto \lambda^{9.5}\frac{v^2}{\LNP^2}$,
and $cc,uu \to sd\propto \lambda^{13}\frac{v^2}{\LNP^2}$, which
is beyond the sensitivity of  the meson decay searches
listed in table \ref{tab:exptme}.

The CKM angles can also enter in the mixing of  the singlet
quark current   $(\overline{d}_p \g_\a d_p)$ into
doublets   $(\overline{q}_i \g_\a q_j)$. Similarly to
the doublet to singlet mixing discussed above,
the effects of CKM are beyond  upcoming
experimental sensitivities. A novel feature in this case
is that approximating the CKM angles to vanish can
generate flavour change when there is none:  an $s_R$
is transformed into  $q_c$ at a Higgs vertex, whose  lower component
matches  at  $m_W$ onto $\sum_p V_{cp} d_{Lp}$.

 Finally,  there are diagrams with one Higgs vertex on the
 quark line and one on a lepton line, which {\it eg} mix vector
 and scalar operators. The  mixing from scalar into
 vector operators, such as 
 ${\cal O}_{LEDQ} \to \{{\cal O}_{LQ}^{(1)}  , {\cal O}_{EQ}\}$
 can be neglected because the lepton Yukawas are smaller
 than  that of the $b$, so any quark-flavour-changing contribution
 is more suppressed than
 the   $(\overline{d}_p \g_\a d_p) \leftrightarrow (\overline{q}_i \g_\a q_j)$
 mixing discussed above.
It is also the case that quark-flavour-changing mixing from vectors
to scalars is below the  sensitivity of upcoming experiments,
despite that the experimental sensitivity to scalar operators
can be better than to vectors (see table \ref{tab:exptme}).
In the case of $\mu \leftrightarrow e$  searches, this is because
 the mixing is suppressed by $y_\mu\sim \lambda^5$, and
 for $\tau \leftrightarrow \ell$ searches,  the experiments are
 less sensitive. 

 So we conclude that CKM angles can be neglected in the SMEFT
 RGEs for LFV operators, provided that one runs in the
 $Y_uY_u^\dagger $ eigenbasis for the $\{q_L\}$, and
 that CKM mixing is retained in matching at $m_W$.

\subsection{LFV Yukawa couplings}
\label{ssec:yuk}

In the SM, the  Yukawa  matrix of the charged
leptons is the  only  basis-choosing interaction in the
leptonic sector --- the gauge interactions are ``universal'',
that is, proportional to  the identity matrix in generation space,
so without eigenvectors. In the  real world (not described
by the SM), the neutrino mass matrix provides another eigenbasis,
but the magnitude of  neutrino  masses is so small
that their  direct GIM-suppressed  contribution to  LFV
is irrelevant (Instead, they provide motivation
search for LFV).

 LFV operators  that are added to  the Lagrangian
 below the weak scale  are inevitably written
 in the mass eigenstate basis of the charged leptons.
Above the weak scale in SMEFT, there are two possibilities:
the  mass eigenstate basis, or the Yukawa eigenstate basis ---
which may be  different in the presence of the  operator
${\cal O}_{eH}$.  The physics, of course, cannot depend on
a basis choice, but the calculation may be more intuitive
and simple in somes bases than  in others.  So which  is
the best choice?

Suppose  one thinks top-down; then at $\LNP$,  the
New Physics model is matched to  the SM +operators.
The obvious basis in this case for SMEFT is the  $D_e$-basis
where the lepton Yukawa matrix is diagonal : $Y_e = D_e =
{\rm diag} \{ y_\tau, y_\mu , y_e\}$. This choice is  motivated
by  LFV  being   a NP effect, 
and ensures  that the SMEFT
RGEs, which describe SM dynamics,  cannot change the flavours of  lepton legs.

However, when the Higgs gets a vev in
the presence of the ${\cal O}_{eH}$ operator,
the  $D_e$ basis may no longer  be the mass eigenstate
basis, due to additional off-diagonal contributions
of  ${\cal O}_{eH}$ to the mass matrix.  So a basis rotation 
during the matching at $m_W$ would be required,
from the $D_e$ basis to the mass eigenstate
basis  in which the restrictive low-energy
constraints are expressed.
Current constraints/sensitivities  on the
off-diagonal elements of ${\cal O}_{eH}$
imply that 
 the angles of  this rotation are small:
 estimating $\theta_{ij} \sim C_{eH}^{ij}v^3/(\LNP^2{\rm max}\{m_i,m_j\})$
 for $i\neq j$ gives
\beq
\theta_{\ell\tau},~ \theta_{ \tau \ell} \lsim \lambda~~~,~~~
\theta_{e \mu} , ~\theta_{\mu e} \lsim \lambda^4
\label{thetas}
\eeq
where $\ell \in\{e,\mu\}$.

If the   New Physics scale is sufficiently high
that only dimension six operators are relevant,
 one might hope to neglect this rotation in
matching, because the angles  are  $\propto Cv^2/\LNP^2$,
so any effect on a NP operator  would be ${\cal O}(1/\LNP^4)$. 
(Below $m_W$, there are also contact interactions
induces by the $W,Z,h$,  which could becomes flavour-changing
under a basis rotation.  However,
the $W$ and
$Z$   interactions are  ``universal'', so unconcerned by
basis rotations, and the higgs-mediated operators
are  suppressed by SM Yukawas, so the dimension six
flavour-changing operators induced by the rotation are unobservable.)
However, as previously discussed, LFV  data can  have sensitivity
to  operators suppressed by  ${\cal O}(1/\LNP^4)$,
and the mixing angles of eqn (\ref{thetas}) are also
enhanced by inverse Yukawas.
The power-counting rules suggest that
flavour-diagonal coefficients at $\LNP \sim 4 $ TeV could
be rotated into $\tau \leftrightarrow \ell$ 
 operators
 suppressed by $\lambda^5$, and into
 $\mu \leftrightarrow e$ suppressed by $\lambda ^8$.
This is within current experimental sensitivities.

We advocate {\it not} making the transformation from the mass to
Yukawa eigenstate basis at $m_W$.  This is because  the rotation is unknown,
and the angles are insufficiently suppressed (see eqn\ref{thetas}).
Instead, we remain in
the mass eigenstate basis above the weak scale; 
this is consistent with our bottom-up perspective,
because it is the basis where the constraints apply. 
The lepton Yukawa matrix can be off-diagonal
in this basis(see eqn \ref{yukawasl}), but the
off-diagonals $\sim \theta_{ij} y_j$ are much smaller than the
$\theta_{ij}$s of eqn (\ref{thetas}) because
they are  suppressed also by small lepton Yukawas. 
The powercounting
suggests that they can be neglected in the RGEs,
for instance
$$
[Y_e]_{\mu e} \frac{\log}{16\pi^2} \frac{v^2}{\LNP^2} \lsim \lambda^{15}~~~.
$$
So in practise, we work in the mass eigenstate basis at
all scales, but treat the lepton Yukawa matrix as diagonal
in the RGEs of SMEFT. The inconvenience  of this choice is that
in matching a model onto the operators, one must
 identify the mass eigenstate basis in the model, and
  obtain operator coefficients in that basis.  

\subsection{ LFV with $\tau$s}
\label{ssec:tau}

This section briefly discusses
the ingredients required for a ``leading order''  
SMEFT study of  LFV among the $\tau$s.

For the majority of $\tau$ LFV processes listed in
Table \ref{tab:exptme} there is sensitivity to Wilson
coefficients that are $\gsim\lambda^5$. Since a loop costs
a factor $\lambda^2$, loop effects in the $\tau$ sector could be
relevant for $(v^2/\Lambda_{\rm NP}^2)\geq \lambda^3$, but
this implies a New Physics scale within the LHC reach.

In the case of the more sensitive $\tau\to e(\mu)\gamma$ searches,
the corresponding diagrams can be power counted  as for  $\mu\to e \gamma$,
replacing the muon leg with a tau leg. Since the constraints
concern dipole coefficients defined with a built-in yukawa
of the heavier lepton, we encounter two possibilities in the diagrams:
\begin{itemize}
	\item either one Higgs leg is attached to the decaying lepton line and the power counting estimate is the same,
	\item or no Higgs-heavy lepton vertex is present and the diagrams are suppressed by a factor $y_\mu/y_\tau=2\lambda^2$ with respect to the corresponding $\mu \to e \gamma$ one.
\end{itemize}
In both cases, given the  lesser sensitivity in the $\tau$ sector,
we can conclude that any approximation that we justify through
power counting 
for $\mu$-s  is also valid for $\tau$ LFV processes.

As a result,  two-loop anomalous dimensions should be irrelevant in
$\tau\leftrightarrow \ell$ processes,  due to  the estimated
suppression $\sim \lambda^5$ of two-loop diagrams.
This should remain true even in the case of $\tau\to e(\mu)\gamma$.

Furthermore, the requirement of eq.~(\ref{eqnn}) on 8-dimensional operator coefficients for $\Lambda_{\rm NP}\gsim4$ T$e$V 
\begin{equation*}
	^{(8)}C\lsim \lambda^8
\end{equation*}
is sufficient to argue that any $\tau$ LFV observable is not sensitive to dimension eight operators.


\section{Summary}
\label{sec:summ}

Effective Field Theory can be envisaged from a bottom-up or
top-down perspective. 
In bottom-up EFT for Lepton Flavour Change(LFV),
the aim is  to  map experimental
  constraints onto the   correct sum of operator
  coefficients at  the New Physics scale $\LNP$, in order to
  identify the area in coefficient
  space where BSM models must sit.
  From a top-down perspective,  one can   map a  LFV model  onto
  operator coefficients at $\LNP$,  calculate
  observables using  EFT, and 
  this  should  correctly reproduce model  predictions
to within a calculable uncertainty.
In both perspectives, 
the EFT  calculation must include correctly  every operator coefficient
that could contribute to an observable,  irrespective
of its dimension or  of the order in the loop or
coupling expansions.

To ensure that  we use SMEFT correctly for
describing LFV,
  we introduced a power-counting scheme,  that allows
  to organise  all the SMEFT  perturbative  expansions --- in loops,  couplings,
  mixing angles and the ratio of the weak scale to  the New Physics
  $v/\LNP$ ---in terms of a  small
  ``Cabbibbo-Wolfenstein-like'' parameter $\lambda\approx 0.2$.
  This power-counting scheme is described
  in section \ref{ssec:lambda}, and summarised in
  table \ref{tab:pc}.
  The future reach of various experiments can be
  expressed  in powers of $\lambda$ (see
   table \ref{tab:exptme}) --- so for instance,
   the upcoming  MEGII experiment searching for  $\meg$ could
   probe dipole coefficients  up to ${\cal O}( \lambda^{12})$.
   Then one can draw diagrams, arising at various orders
   in the different perturbative expansions, and do
   two things; first, compare different contributions of
  an operator to an observable, to identify the leading one,
(see eg section \ref{ssec:implement} and \ref{ssec:2loop}).
And secondly,    one can
    determine which operators can affect
   which observables by comparing the power-counting estimates  
   to the future experimental sensitivity.  Some examples are given
   in     Section \ref{ssec:implement}.

For LFV operators, the SMEFT expansion in
operator dimension can be written
as an expansion in $v^2/\LNP^2$, where
the New Physics scale $\LNP$  
  plays two roles in our manuscript.
One one hand, it
is the  unknown mass of the lightest lepton flavour changing new particle
(see the Lagrangian of eqn (\ref{L})),
which  we take   ``beyond
the reach of the LHC'': $\LNP\gsim 4 $ TeV
(so $v^2/\LNP^2 \lsim {\cal O}(\lambda^4)$ in the powercounting scheme).
However,  since $\LNP$
is unknown, we simultaneously
count the order of an operators contribution
by the scale it could probe
with a coefficient  of  ${\cal O}(1/\LNP^{2n})$.

In the  SMEFT,  there are already many  operators
at  dimension six, and their  RGEs are only known
at one-loop.   So  in section \ref{sec:qestions}, 
we  use the powercounting scheme to
explore  whether  dimension six operators
and one-loop RGEs
 are sufficient
to describe LFV at the sensitivity of
experiments under construction. 
Section  \ref{ssec:2loop}  suggests that some 
two-loop anomalous dimensions are
required for $\mu\leftrightarrow e$ flavour
change,  when  $  \LNP \lsim 20$ TeV.
The calculation
of these anomalous dimensions  is
in progress \cite{SJL}.

Section \ref{ssec:dim8}   finds that
upcoming  $\mu\leftrightarrow e$  data
can be sensitive to  dimension
eight SMEFT operators, about four dozen
of them for $ \LNP\gsim 4$ TeV,
but none at scales $ \LNP\gsim 100$ TeV.
The relevant dimension eight operators
match onto three-or four-point interactions below
the weak scale, and can be divided
into   two sets:
 those which are the lowest-dimension
SMEFT operator inducing
a given contact interaction below $m_W$,
and a second set that induces
low-energy contact interactions already
present   at dimension six. 
The scale $\LNP$ up to which  the
operators can be  relevant is given in
 tables  \ref{tab:d8LNP>}
and   \ref{tab:d8LNP>2}.
These dimension  eight operators are listed
in Appendix \ref{app:O8}, and
are included in the matching onto
operators below $m_W$ in Appendix \ref{app:match}.


The power counting scheme can also be
used to simplify and   streamline  calculations
with the  existing  SMEFT operators
and RGEs, for instance by
neglecting  flavour-changing SM interactions.
We perform two such exercises;
section \ref{ssec:CKM} checks that CKM
mixing can be neglected  in the RGEs for LFV
operators, provided that it is included
in matching, and  that the SMEFT  RGEs run in the
$Y_uY_u^\dagger$ eigenbasis for  the $\{q_L\}$.
Section \ref{ssec:yuk} explores the case
where operators of the form
$C^{ij}(H^\dagger H)^n \overline{\ell}_iHe_j$,
with $i\neq j$,  are allowed to contribute
to the charged  lepton mass matrix.
This implies that  in the charged lepton
mass eigenstate basis (where all experimental
constraints are given), the charged
lepton Yukawa $Y_e$  has  unknown off-diagonal elments.
The power-counting suggests  that
if these flavour-changing Yukawas are
below current experimental sensitivities, they
can be neglected in the SMEFT RGEs.

In this manuscript, we estimated lower bounds on the  scale
$\Lambda_{NP}$, such that  the predictions
of  lepton flavour changing New Physics  models
from beyond $\Lambda_{NP}$  can be  obtained  with the dimension six
operators of SMEFT and their one-loop RGEs. 
These results could be used to
motivate, or justify, SMEFT studies of LFV.
It could be interesting to  perform
a similar study in  the EFT with  a ``non-linear
realisation'' of the Higgs sector\cite{BC,P},
and also to perform a
more systematic  expansion  to ensure
that the leading terms  are identified.


\subsection*{Acknowledgements}
We thank Junji Hisano for  proposing the initial powercounting scheme,
and Luca Silvestrini for useful conversations.

\appendix


\section{Some LFV Operators of dimension eight}
\label{app:O8}
Section \ref{ssec:dim8} showed that $\mu\leftrightarrow e$ processes can be sensitive to some SMEFT operators of dimension eight, if these have {\cal O}(1) coefficients at $\LNP\gsim 4 $ TeV.  This appendix  lists  the relevant
 operators, following the notation of \cite{Murphy:2020rsh}.

The LFV operators given here  are required to  match onto
low energy  operators involved in the processes
of  Table \ref{tab:exptme}, so  derivative operators,
and those involving more than four particles at low energy,
are neglected. In addition, operators of the
form  $ \mu_H^2\times$ dimension six, where  $\mu_H^2$
is the  Higgs mass$^2$ term  in the Lagrangian,
are neglected because  in matching onto operators
below $m_W$, the potential minimisation condition
relates  $\mu_H^2$  to $H^\dagger H$. Furthermore, we restrict our list to operators that are $\mu \leftrightarrow e$ flavour changing but flavour diagonal in the two other fermion legs, as the low energy observables constrain operator with this flavour structure.

The four-fermion operators of dimension eight can be obtained by adding two Higgs
fields to dimension six four-fermion operators, or by multiplying two renormalizable Lagrangian terms.
Dimension six operators can be multiplied by the singlet product ($H^\dagger H$), but the Higgses can also contract with specific doublets; when the Higgs gets a vev,  this feature
induces a low-energy operator involving only some
SU(2) partners.
For instance, the dimension eight operator
$$
(\overline{\ell}_\a \tilde{H} \gamma_\rho\tilde{H}^\dagger \ell_\b)
(\overline{q}\gamma^\rho q) \to
(\overline{\nu}_\a  \gamma_\rho \nu_\b)
(\overline{u}\gamma^\rho u + \overline{d}\gamma^\rho d) ~~~.
$$
This operator induces ``Non-Standard neutrino Interactions''\cite{NSI}, which
can be searched for at neutrino experiments, without 
inducing tree-level flavour-change among charged leptons. Exploiting $\text{SU}(2)$ identities, these operators can be expressed as linear combinations of dim6$\times(H^\dagger H)$ and the following operator
$$
(\overline{\ell}_\a \tau^I \gamma_\rho \ell_\b)
(\overline{q}\gamma^\rho q)(H^\dagger \tau^I H).
$$
Adopting the convention of \cite{Murphy:2020rsh}, we retain the triplet contractions in the operator basis. Since we are interested in the contribution of dimension eight operators to LFV observables, we organize the operator list according to whether a dimension six version exists or does not exist.

We display operators with ``standard'' flavour indices and we don't include the permutations that will be matched to the same low energy interaction, as discussed in Appendix \ref{app:match}. 
\subsection{Dimension eight not present at dimension six}
\subsubsection{Four-fermion}
SU(2) invariance and its chiral nature forbid SMEFT dimension six counterparts of some four-fermion contact interaction of the QCD$*$QED invariant Lagrangian, forcing their appearance at dimension eight. In the case of four-fermion operators with four-lepton legs these are the tensor operators
\begin{align*}
	\mathcal{O}^{(4)e\mu kk}_{L^2 E^2H^2}=(\bar{l}_eH\sigma^{\alpha\beta} e_\mu)(\bar{l}_kH\sigma_{\alpha\beta} e_k) 
\end{align*}
where $k \in\{ e,\mu,\tau\}$.
They can be related to the scalars $\mathcal{O}^{(3)ijkl}_{L^2 E^2H^2}=(\bar{l}_iH e_j)(\bar{l}_kH e_l)$ of the basis \cite{Murphy:2020rsh} thanks to the following Fierz identity
$$
\mathcal{O}_{L^2 E^2H^2}^{(4)e\mu kk}=-8\mathcal{O}_{L^2 E^2H^2}^{(3)ekk\mu}-4\mathcal{O}_{L^2 E^2H^2}^{(3)e\mu kk}.
$$
Given that the tensors mix with the dipole, we retain both operators in the matching conditions of Appendix \ref{app:match}, keeping in mind that we can remove the redundancy by means of the above identity. 

For four-fermion interaction involving two-lepton and two-quark legs, the dimension eight operators that do not arise at dimension six are
\begin{align*}	
	\mathcal{O}^{(3)e\mu nn}_{LEDQH^2}&=(\bar{\ell}_e He_\mu)(\bar{q}_n H d_n)\qquad 
	\mathcal{O}^{(4)e\mu nn}_{LEDQH^2}=(\bar{\ell}_e \sigma^{\alpha\beta}He_\mu)(\bar{q}_n \sigma_{\alpha\beta}H d_n)\\
	\mathcal{O}^{(5)e\mu nn}_{LEQUH^2}&=(\bar{\ell}_eHe_\mu)(\bar{u}_n\tilde{H}^\dagger q_n).
\end{align*}
where $n $ is a quark generation index. 
In this case, the scalar and tensor operator for down-type quarks are independent and cannot be related by means of Fierz identities.
\subsubsection{Two-lepton operators}
Two-lepton and two-gauge boson operators firstly appear at dimension eight
\begin{eqnarray}
	\mathcal{O}^{(1)e\mu}_{LEG^2H}&=&(\bar{\ell}_eH e_\mu)G^A_{\alpha\beta}G^{A\alpha\beta}\qquad 	\mathcal{O}^{(2)e\mu}_{LEG^2H}=(\bar{\ell}_eH e_\mu)G^A_{\alpha\beta}\tilde{G}^{A\alpha\beta} \nonumber\\
	\mathcal{O}^{(1)e\mu}_{LEW^2H}&=&(\bar{\ell}_eH e_\mu)W^I_{\alpha\beta}W^{I\alpha\beta}\qquad 	\mathcal{O}^{(2)e\mu}_{LEW^2H}=(\bar{\ell}_eH e_\mu)W^I_{\alpha\beta}\tilde{W}^{I\alpha\beta} \nonumber\\
	\mathcal{O}^{(1)e\mu}_{LEB^2H}&=&(\bar{\ell}_eH e_\mu)B_{\alpha\beta}B^{\alpha\beta}\qquad 	\mathcal{O}^{(2)e\mu}_{LEB^2H}=(\bar{\ell}_eH e_\mu)B_{\alpha\beta}\tilde{B}^{\alpha\beta} \nonumber\\
	\mathcal{O}^{(1)e\mu}_{LEWBH}&=&(\bar{\ell}_e\tau^IH e_\mu)B^{\alpha\beta}W^I_{\alpha\beta}\qquad 	\mathcal{O}^{(2)e\mu}_{LEWBH}=(\bar{\ell}_e\tau^I H e_\mu)B_{\alpha\beta}\tilde{W}^{I\alpha\beta} \nonumber
\end{eqnarray}
and provide the leading order matching contribution to the dimension seven  two-photon $\mathcal{O}_{FF,Y}=(\bar{e}P_Y\mu)F_{\alpha\beta}F^{\alpha\beta}$,$\mathcal{O}_{F\tilde{F},Y}=(\bar{e}P_Y\mu)F_{\alpha\beta}\tilde{F}^{\alpha\beta}$ and two-gluon $\mathcal{O}_{GG,Y}=(\bar{e}P_Y\mu)G^A_{\alpha\beta}G^{A\alpha\beta}$, $\mathcal{O}_{G\tilde{G},Y}=(\bar{e}P_Y\mu)G^A_{\alpha\beta}\tilde{G}^{A\alpha\beta}$ operators of the low energy Lagrangian, whose coefficients are constrained by searches of $\mu \to e$ conversion in nuclei. 
\subsection{Dimension eight operators present at dimension six}
\subsubsection{Four-fermion}
The four-fermion operators with four lepton legs that also appear at dimension six are
\begin{align*}
	\mathcal{O}^{(1)e\mu kk}_{L^4H^2}&=(\bar{\ell}_e\gamma^\alpha \ell_\mu)(\bar{\ell}_k\gamma_\alpha \ell_k)(H^\dagger H)\qquad 	\mathcal{O}^{(2)e\mu kk}_{L^4 H^2}=(\bar{\ell}_e\gamma^\alpha \ell_\mu)(\bar{\ell}_k\tau^I\gamma_\alpha \ell_k)(H^\dagger\tau^I H)\\
	\mathcal{O}^{(1)e\mu kk}_{L^2 E^2H^2}&=(\bar{\ell}_e\gamma^\alpha \ell_\mu)(\bar{e}_k\gamma_\alpha e_k)(H^\dagger H)\qquad 		\mathcal{O}^{(2)e\mu kk}_{L^2 E^2H^2}=(\bar{\ell}_e\tau^I\gamma^\alpha \ell_\mu)(\bar{e}_k\gamma_\alpha e_k)(H^\dagger \tau^IH)\\
	\mathcal{O}^{e\mu kk}_{E^4H^2}&=(\bar{e}_e\gamma^\alpha e_\mu)(\bar{e}_k\gamma_\alpha e_k)(H^\dagger H),
\end{align*}
where $k=e,\mu, \tau$. 

In addition, the four-fermion operators containing two-lepton and two-quark legs are:
\begin{align*}
	\mathcal{O}^{(1)e\mu nn}_{L^2Q^2H^2}&=(\bar{\ell}_e\gamma^\alpha \ell_\mu)(\bar{q}_n\gamma_\alpha q_n)(H^\dagger H)\qquad 	\mathcal{O}^{(2)e\mu nn}_{L^2Q^2H^2}=(\bar{\ell}_e\tau^I\gamma^\alpha \ell_\mu)(\bar{q}_n\gamma_\alpha q_n)(H^\dagger\tau^I H)\\
	\mathcal{O}^{(3)e\mu nn}_{L^2Q^2H^2}&=(\bar{\ell}_e\tau^I\gamma^\alpha \ell_\mu)(\bar{q}_n \tau^I\gamma_\alpha q_n)(H^\dagger H)\qquad 	\mathcal{O}^{(4)e\mu nn}_{L^2Q^2H^2}=(\bar{\ell}_e\gamma^\mu \ell_\mu)(\bar{q}_n\tau^I\gamma_\mu q_n)(H^\dagger\tau^I H)\\
	\mathcal{O}^{(5)e\mu nn}_{L^2Q^2H^2}&=\epsilon^{IJK}(\bar{\ell}_e\tau^I\gamma^\mu \ell_\mu)(\bar{q}_n\tau^J\gamma_\mu q_n)(H^\dagger \tau^K H)\qquad 		\mathcal{O}^{(1)e\mu nn}_{L^2U^2H^2}=(\bar{\ell}_e\gamma^\alpha \ell_\mu)(\bar{u}_n\gamma_\mu u_n)(H^\dagger H)\\
	\mathcal{O}^{(2)e\mu nn}_{L^2U^2H^2}&=(\bar{\ell}_e\tau^I\gamma^\alpha \ell_\mu)(\bar{u}_k\gamma_\alpha u_l)(H^\dagger \tau^I H)\qquad \mathcal{O}^{(1)e\mu nn}_{L^2D^2H^2}=(\bar{\ell}_e\gamma^\alpha \ell_\mu)(\bar{d}_k\gamma_\alpha d_l)(H^\dagger H)\\
	\mathcal{O}^{(2)e\mu nn}_{L^2D^2H^2}&=(\bar{\ell}_e\tau^I\gamma^\alpha \ell_\mu)(\bar{d}_n\gamma_\alpha d_n)(H^\dagger \tau^I H)\qquad \mathcal{O}^{(1)e\mu nn}_{E^2Q^2H^2}=(\bar{e}_e\gamma^\alpha e_\mu)(\bar{q}_n\gamma_\alpha q_n)(H^\dagger H)\\
	\mathcal{O}^{(2)e\mu nn}_{E^2Q^2H^2}&=(\bar{e}_e\gamma^\alpha e_\mu)(\bar{q}_n\tau^I\gamma_\alpha q_n)(H^\dagger \tau^I H)\qquad \mathcal{O}^{e\mu nn}_{E^2U^2H^2}=(\bar{e}_e\gamma^\alpha e_\mu)(\bar{u}_n\gamma_\alpha u_n)(H^\dagger H)\\
	\mathcal{O}^{e\mu nn}_{E^2D^2H^2}&=(\bar{e}_e\gamma^\alpha e_\mu)(\bar{d}_n\gamma_\alpha d_n)(H^\dagger H)\qquad 
	\mathcal{O}^{(1)e\mu nn}_{LEDQH^2}=(\bar{\ell}_e e_\mu)(\bar{d}_n q_n)(H^\dagger H)\\	\mathcal{O}^{(2)e\mu nn}_{LEDQH^2}&=(\bar{\ell}_e e_\mu)\tau^I(\bar{d}_n q_n)(H^\dagger\tau^I H)\qquad
	\mathcal{O}^{(1)e\mu nn}_{LEQUH^2}=(\bar{\ell}_e e_\mu)\epsilon(\bar{q}_n u_n)(H^\dagger H)\\ 
	\mathcal{O}^{(2)e\mu nn}_{LEQUH^2}&=(\bar{\ell}_e e_\mu)\tau^I\epsilon(\bar{q}_n u_n)(H^\dagger \tau^I H)\qquad
	\mathcal{O}^{(3)e\mu nn}_{LEQUH^2}=(\bar{\ell}_e\sigma^{\alpha\beta} e_\mu)\epsilon(\bar{q}_n \sigma_{\alpha\beta}u_n)(H^\dagger H)\\ 
	\mathcal{O}^{(4)e\mu nn}_{LEQUH^2}&=(\bar{\ell}_e\sigma^{\alpha\beta} e_j)\tau^I\epsilon(\bar{q}_n \sigma_{\alpha\beta}u_n)(H^\dagger\tau^I H)
\end{align*}
where $n=1,2,3$ runs over the quark generation space. 
\subsubsection{Two-lepton operators}
Two-lepton operators include the eight dimensional dipoles
\begin{align*}
	\mathcal{O}^{(1)e\mu}_{LEWH^3}&=(\bar{\ell}_e \tau^I H\sigma^{\alpha\beta} e_\mu)W^I_{\alpha\beta}(H^\dagger H)\\
	\mathcal{O}^{(2)e\mu}_{LEWH^3}&=(\bar{\ell}_e H\sigma^{\alpha\beta} e_\mu)W^I_{\alpha\beta}(H^\dagger\tau^I H)\\
	\mathcal{O}^{e\mu}_{LEBH^3}&=(\bar{\ell}_i H\sigma^{\alpha\beta} e_j)B_{\alpha\beta}(H^\dagger H)
\end{align*}
and the following operators
\begin{align*}
	\mathcal{O}^{(1)e\mu}_{L^2H^4D}&=i(\bar{\ell}_e \gamma^\alpha \ell_\mu)(H^\dagger\overset\leftrightarrow{D}_\alpha H)(H^\dagger H)\qquad 
	\mathcal{O}^{(2)e\mu}_{L^2H^4D}=i(\bar{\ell}_e \tau^I\gamma^\alpha \ell_\mu)[(H^\dagger\overset{\leftrightarrow}{D_{\alpha} ^I} H)(H^\dagger H)+(H^\dagger\overset\leftrightarrow{D}_{\alpha}  H)(H^\dagger\tau^I H)]\\ 
	\mathcal{O}^{e\mu}_{E^2H^4D}&=i(\bar{e}_e \gamma^\alpha e_\mu)(H^\dagger\overset{\leftrightarrow}D_{\alpha}  H)(H^\dagger H)\qquad
	\mathcal{O}^{e\mu}_{LEH^5}=(\bar{\ell}_e He_\mu)(H^\dagger H)^2,
\end{align*}
where 
\begin{align*}
	iH^\dagger\overset\leftrightarrow{D}_\mu H&\equiv iH^\dagger (D_\mu H)-i(D_\mu H^\dagger)H\\
	iH^\dagger\overset{\leftrightarrow}{D_{\mu} ^I} H&\equiv iH^\dagger \tau^I(D_\mu H)-i(D_\mu H^\dagger)\tau^I H.
\end{align*}
Following Electroweak Spontaneous Symmetry Breaking, the second set of operators are matched onto four fermion contact interactions at low energy, after integrating out the heavy $Z$, $h$ bosons at $m_W$.

\section{Tree matching at $m_W$   with LFV operators to dimension eight}
\label{app:match}

This section  presents the tree level matching conditions at $m_W$
of   $\mu\leftrightarrow e$ flavour-changing   SMEFT operators,
including the  dimension eight
operators listed in the previous section.  The operator basis
below $m_W$ is given in the notation of \cite{C+C,megmW}.

\subsection{Dipoles and 2 photon(gluon)}
Below $m_W$, there  are the dipole operators of
two chiralities, and  operators with two photons
or two gluons.
Above $m_W$,
there is a  dimension six   dipole  operator
for hypercharge,  and another  one for SU(2). 

Since the photon is the combination $A_\mu= \cos\theta_W B_\mu+\sin\theta_W W_\mu^3\equiv c_W B_\mu+s_W W_\mu^3$, the low energy dipole coefficient (on the left) is matched onto the dimension six and eight SMEFT dipoles (on the right) as
\bea
C^{e\mu}_{D,R} = c_W \left(C^{e\mu}_{EB}+\frac{v^2}{y_\mu\LNP^2}C^{e\mu}_{LEBH^3}\right) -s_W  \left[C^{e\mu}_{EW}+\frac{v^2}{y_\mu\LNP^2}\left(C^{e\mu}_{LEWH^3(1)}+C^{e\mu}_{LEWH^3(2)}\right)\right]\\
C^{e\mu}_{D,L} = c_W \left(C^{\mu e*}_{EB}+\frac{v^2}{y_\mu\LNP^2}C^{\mu e*}_{LEBH^3}\right) -s_W  \left[C^{\mu e*}_{EW}+\frac{v^2}{y_\mu\LNP^2}\left(C^{\mu e*}_{LEWH^3(1)}+C^{\mu e*}_{LEWH^3(2)}\right)\right]
\label{Dmatch}
\eea
where the $-$ sign is due to the $\tau^3$ matrix.
In addition, since matching ``at tree level'' mean tree-level in the low-energy theory,  loop diagrams in the theory above $m_W$  composed of  heavy particles
can be included. We follow \cite{megmW} (see\cite{DS} for a more recent calculation), and  retain the two-loop Barr-Zee diagrams, in which a Higgs leg connect a $W$ or $t$ loop with the neutral Higgs flavour changing vertex of eq.~(\ref{appFR:Y1}), and the one loop $Z-$exchange diagram where one $Z$ vertex is flavour changing. The former give the matching condition
\beq
\Delta C^{e\mu}_{D,L}(m_W)  \simeq -
C^{\mu e *}_{EH}(m_W)  
\left[ 
\frac{e\alpha }{16\pi^3y_\mu } \left(  Q_t^2   N_c Y^2_t
-\frac{7}{2}   \right)
\right]  \simeq  C^{\mu e *}_{EH}(m_W)  \left[ 
\frac{e\alpha }{8\pi^3y_\mu } 
\right],
\eeq
while the latter give
 \bea
\Delta C^{e \mu }_{D,L } (m_W)  &\simeq& \frac{e }{16\pi^2}
g^e_L C^{e \mu}_{HE} (m_W) \nonumber\\
\Delta C^{e \mu  }_{D,R} (m_W)  &\simeq& \frac{e }{16\pi^2}
g^e_R \left( C^{e \mu}_{HL(1)}  (m_W)
+  C^{e \mu}_{HL(3)} (m_W)
\right),
\eea
where $g_L^e, g_R^e$ are defined in the Feynman rule for $Z$ couplings to leptons
$-i \frac{g}{2 c_W}(g^e_L P_L + g^e_R P_R)$ as
$ g_R^e = 2 s_W^2$, and 
$g_L^e = -1+2 s_W^2$.

For the 2 photon and 2 gluon operators the matching conditions are
\begin{eqnarray}
	C^{e\mu}_{FF,R}&=&\frac{v}{\LNP}\left(c_W^2 C^{e\mu}_{LEB^2H(1)}-s_Wc_W C^{e\mu}_{LEWBH(1)}+s_W^2 C^{e\mu}_{LEW^2H(1)}\right)\\
	C^{e\mu}_{FF,L}&=&\frac{v}{\LNP}\left(c_W^2 C^{\mu e*}_{LEB^2H(1)}-s_Wc_W C^{\mu e*}_{LEWBH(1)}+s_W^2 C^{\mu e*}_{LEW^2H(1)}\right)\\
	C^{e\mu}_{F\tilde{F},R}&=&\frac{v}{\LNP}\left(c_W^2 C^{e\mu}_{LEB^2H(2)}-s_Wc_W C^{e\mu}_{LEWBH(2)}+s_W^2 C^{e\mu}_{LEW^2H(2)}\right)\\
	C^{e\mu}_{F\tilde{F},L}&=&\frac{v}{\LNP}\left(c_W^2 C^{\mu e*}_{LEB^2H(2)}-s_Wc_W C^{\mu e*}_{LEWBH(2)}+s_W^2 C^{\mu e*}_{LEW^2H(2)}\right)\\
	C^{e\mu}_{GG,R}&=&\frac{v}{\LNP}C^{e\mu}_{LEG^2H(1)}\qquad C^{e\mu}_{GG,L}=\frac{v}{\LNP}C^{\mu e*}_{LEG^2H(1)}\\
	C^{e\mu}_{G\tilde{G},R}&=&\frac{v}{\LNP}C^{e\mu}_{LEG^2H(2)}\qquad C^{e\mu}_{G\tilde{G},L}=\frac{v}{\LNP}C^{\mu e*}_{LEG^2H(2)}
\end{eqnarray}
\subsection{4-Lepton}\label{4leptonmatch}
Four fermion operators are matched onto the low energy analogue after ESWB and 2 lepton operators are matched after integrating out the heavy $h,W,Z$ bosons . 

The SMEFT operators $\mathcal{O}_{EH}$ and $\mathcal{O}_{LEH^5}$ contribute to the charged leptons mass as 
\begin{equation}
[m_e]^{ij} = v\left([Y_e]^{ij} - C_{EH}^{ij}\frac{v^2}{\LNP^2}- C_{LEH^5}^{ij}\frac{v^4}{\LNP^4}\right).
\end{equation}
while the neutral Higgs Yukawa couplings are
\begin{equation}
\frac{h}{\sqrt{2}}\bar{e}^iP_Re^j\left([Y_e]^{ij} - 3C_{EH}^{ij}\frac{v^2}{\LNP^2}- 5C_{LEH^5}^{ij}\frac{v^4}{\LNP^4}\right)=\frac{h}{\sqrt{2}}\bar{e}^iP_Re^j\left([m_e]^{ij} - 2C_{EH}^{ij}\frac{v^2}{\LNP^2}- 4C_{LEH^5}^{ij}\frac{v^4}{\LNP^4}\right),
\end{equation}
so that in the mass basis for the charged leptons we find a neutral Higgs flavour changing vertex, with the Feynman Rule
\begin{equation}
-i\sqrt{2}\bar{e}^iP_Re^j\left( C_{EH}^{ij}\frac{v^2}{\LNP^2}+ 2C_{LEH^5}^{ij}\frac{v^4}{\LNP^4}\right)
\end{equation}
In SMEFT there are more distinct flavour structures which are matched into the same low energy operators: for example ${\cal O}^{e \mu ff}_{L L} $,
${\cal O}^{ff e \mu}_{L L} $ ,  ${\cal O}^{ f \mu e f}_{L L} $  and 
${\cal O}^{e ff \mu}_{L L} $ all  match onto
the below-$m_W$  LFV operator ${\cal O}^{e \mu ff}_{L L} $. In the following,
we  suppress these permutations for brevity, and write
$$
C^{e \mu ff}_{\rm low~energy} =  C^{e \mu ff}_{\rm SMEFT} +{\rm perm.}
$$
to indicate  that these different flavour structures are
to be summed on the right side of  the matching conditions. These are:
\bea 
C^{e \mu \ell\ell}_{V,RR}& =& C^{e \mu \ell \ell}_{EE}  +C^{e \mu}_{HE}  g_R^e+\frac{v^2}{\LNP^2}\left(C^{e\mu \ell\ell}_{E^4H^2}+C^{e\mu}_{E^2H^4D}g^e_R\right)+\text{perm.}
\label{19}\\%
C^{e \mu \ell \ell}_{V,LR} &=& C^{e \mu \ell \ell}_{LE} 
+ (  C^{e \mu}_{HL,3}
+ C^{e \mu}_{HL,1})g_R^e+\frac{v^2}{\LNP^2}\left[C^{e\mu ll}_{L^2E^2H^2(1)}+C^{e\mu ll}_{L^2E^2H^2(2)}+\left(C^{e\mu}_{L^2H^4D(1)}+2C^{e\mu}_{L^2H^4D(2)}\right)g^e_R\right]
\label{20}  \\
C^{e \mu \ell \ell}_{V,RL}& =& C^{\ell \ell e \mu }_{LE} +  C^{e \mu}_{HE} g_L^e+\frac{v^2}{\LNP^2}\left[C^{\ell \ell e \mu}_{L^2E^2H^2(1)}+C^{\ell \ell e \mu}_{L^2E^2H^2(2)}+C^{e\mu}_{E^2H^4D}g^e_L\right] \label{21}\\
C^{e \mu \ell \ell }_{V,LL} & =&  C^{e \mu \ell \ell}_{LL}
+(  C^{e \mu}_{HL,3}
+ C^{e \mu}_{HL,1}) g_L^e+\frac{v^2}{\LNP^2}\left[C^{e \mu\ell \ell }_{L^4H^2(1)}+C^{e \mu\ell \ell }_{L^4H^2(2)}+\left(C^{e\mu}_{L^2H^4D(1)}+C^{e\mu}_{L^2H^4D(2)}\right)g^e_L\right]+{\text{perm.}}
\label{23}\\
&&\nonumber\\
C^{e \mu \ell \ell}_{S,RR}& =&  - \frac{m_\ell C^{e \mu  }_{EH} v}{m_h^2}+\frac{v^2}{\LNP^2}\left(C^{e\mu\ell\ell}_{L^2E^2H^2(3)}- 2\frac{m_\ell C^{e \mu  }_{LEH^5} v}{m_h^2}\right)+\text{perm.}\label{24} \\
C^{e \mu \tau \tau}_{S,LR}& =& -2 C_{LE}^{\tau \mu e \tau} - \frac{m_\tau C^{ \mu e *}_{EH} v}{m_h^2}-\frac{v^2}{\LNP^2}\left[2\left(C^{\tau \mu e \tau}_{L^2E^2H^2(1)}+C^{\tau \mu e \tau}_{L^2E^2H^2(2)}\right)+2\frac{m_\tau C^{ \mu e *}_{LEH^5} v}{m_h^2}\right]\\
C^{e \mu \tau \tau}_{S,RL}& =& -2 C_{LE}^{e \tau \tau \mu}
- \frac{m_\tau C^{e \mu }_{EH} v}{m_h^2}-\frac{v^2}{\LNP^2}\left[2\left(C^{e \tau \tau \mu}_{L^2E^2H^2(1)}+C^{e \tau \tau \mu}_{L^2E^2H^2(2)}\right)+2\frac{m_\tau C^{ e\mu }_{LEH^5} v}{m_h^2}\right] \\
C^{e \mu \ell \ell}_{S,LL}& =&  - \frac{m_\ell C^{ \mu e*  }_{EH} v}{m_h^2}+\frac{v^2}{\LNP^2}\left(C^{\mu e \ell\ell*}_{L^2E^2H^2(3)}-2\frac{m_\ell C^{\mu e*  }_{LEH^5} v}{m_h^2}\right)+\text{perm.}
\label{27}\\
&& \nonumber\\
C^{e \mu \tau \tau}_{T,RR} &= & \frac{v^2}{\LNP^2}C^{e\mu\tau\tau}_{L^2E^2H^2(4)} 
\label{28} \\
C^{e \mu \tau \tau}_{T,LL} &= & \frac{v^2}{\LNP^2}C^{\mu e\tau\tau*}_{L^2E^2H^2(4)}  \label{29}
\eea
where $\ell \in \{e,\mu,\tau \}$.
We see that lepton tensors are matched at tree level only at dimension eight,
and  also that dimension eight operators could be significant for
LL or RR scalars, where the dimension six contribution is Yukawa-suppressed.

\subsection{2 Lepton 2 Quark}
Given that the low energy constraints are expressed in the quark mass eigenstate basis, in the ``bottom up'' approach adopted here, the CKM matrix will act on SMEFT operator coefficients in the matching conditions. As we work in the $u_L-$basis, a CKM weighted sum will appear in matching $d_L$ operators.

Tree-level matching SMEFT dimension six and eight coefficients (on the right) onto low energy coefficients (on the left) results in:
\bea
C^{e \mu u_n u_n}_{LL}& =&
C^{e \mu nn}_{LQ (1)} -  C^{e \mu nn}_{LQ (3)} + g^u_L
( C^{e \mu }_{HL  (1)} + C^{e \mu }_{HL  (3)})  \nonumber \\ &+&\frac{v^2}{\LNP^2}\left[C^{e\mu nn}_{L^2Q^2H^2(1)}+C^{e\mu nn}_{L^2Q^2H^2(2)}-C^{e\mu nn}_{L^2Q^2H^2(3)}-C^{e\mu nn}_{L^2Q^2H^2(4)}+\left(C^{e\mu}_{L^2H^4D(1)}+2C^{e\mu}_{L^2H^4D(2)}\right)g_L^u\right] \\
C^{e \mu d_n d_n}_{LL}& =& 
\sum_{jk} V_{jn} V^*_{kn} (
C^{e \mu jk}_{LQ (1)} +  C^{e \mu jk}_{LQ (3)}) 
+ g^d_L ( C^{e \mu }_{HL  (1)} + C^{e \mu }_{HL  (3)} ) \nonumber \\ 
&+&\frac{v^2}{\LNP^2}\bigg[\sum_{jk} V_{jn} V^*_{kn}\left(C^{e\mu jk}_{L^2Q^2H^2(1)}+C^{e\mu jk}_{L^2Q^2H^2(2)}+C^{e\mu jk}_{L^2Q^2H^2(3)}+C^{e\mu jk}_{L^2Q^2H^2(4)}\right)\\
&+&\left(C^{e\mu}_{L^2H^4D(1)}+2C^{e\mu}_{L^2H^4D(2)}\right)g_L^d\bigg]
\\
C^{e \mu u_n u_n}_{RR}& =& C^{e \mu nn}_{EU} + g^u_R  C^{e \mu }_{HE}+\frac{v^2}{\LNP^2}\left(C^{e \mu nn}_{E^2U^2H^2}+C^{e \mu}_{E^2H^4D}g^u_R\right)\\
C^{e \mu d_n d_n}_{RR}& =& C^{e \mu nn}_{ED} + g^d_R  C^{e \mu }_{HE}+\frac{v^2}{\LNP^2}\left(C^{e \mu nn}_{E^2D^2H^2}+C^{e \mu}_{E^2H^4D}g^d_R\right)\\
C^{e \mu u_n u_n}_{LR}& =& C^{e \mu nn}_{LU}  + g^u_R ( C^{e \mu }_{HL  (1)} + C^{e \mu }_{HL  (3)} )\nonumber \\
&+&\frac{v^2}{\LNP^2}\left[C^{e \mu nn}_{L^2U^2H^2(1)}+C^{e \mu nn}_{L^2U^2H^2(2)}+\left(C^{e\mu}_{L^2H^4D(1)}+2C^{e\mu}_{L^2H^4D(2)}\right)g_R^u\right] \\
C^{e \mu d_n d_n}_{LR}& =&  C^{e  \mu  nn}_{LD}+ g^d_R ( C^{e \mu }_{HL  (1)} + C^{e \mu }_{HL  (3)} )\nonumber \\
&+&\frac{v^2}{\LNP^2}\left[C^{e \mu nn}_{L^2D^2H^2(1)}+C^{e \mu nn}_{L^2D^2H^2(2)}+\left(C^{e\mu}_{L^2H^4D(1)}+2C^{e\mu}_{L^2H^4D(2)}\right)g_R^d\right]\\
C^{e \mu u_n u_n}_{RL} 
& =& C^{e \mu nn}_{EQ}  + g^u_L  C^{e \mu }_{HE}+\frac{v^2}{\LNP^2}\left[C^{e \mu nn}_{E^2Q^2(1)}-C^{e \mu nn}_{E^2Q^2(2)}+C^{e \mu}_{E^2H^4D}g^u_L\right]\\
C^{e \mu d_n d_n}_{RL}& =& 
\sum_{jk} V_{jn} V^*_{kn} 
C^{e \mu jk}_{EQ}  + g^d_L  C^{e \mu }_{HE}+\frac{v^2}{\LNP^2}\left[\sum_{jk} V_{jn} V^*_{kn}\left(C^{e \mu jk}_{E^2Q^2(1)}+C^{e \mu jk}_{E^2Q^2(2)}\right)+C^{e \mu}_{E^2H^4D}g^d_L\right]\\
&&\nonumber
\eea 
\bea
C^{e \mu u_nu_n}_{S,LL}& =& -C_{LEQU}^{* \mu e nn} -  \frac{m_{u_n} v}{m_h^2}  C^{ \mu  e *}_{EH}-\frac{v^2}{\LNP^2}\left(C^{\mu e n n*}_{LEQUH^2(1)}+C^{\mu e n n*}_{LEQUH^2(2)}+2\frac{m_{u_n} v}{m_h^2}  C^{ \mu  e *}_{LEH^5}\right) \label{39} \\
C^{e \mu d_n d_n}_{S,LL}& =& -  \frac{m_{d_n} v}{m_h^2}  C^{ \mu  e *}_{EH}+\frac{v^2}{\LNP^2}\left(\sum_{j}V^*_{jn}C^{\mu e j n*}_{LEQDH^2(3)}-2\frac{m_{d_n} v}{m_h^2}  C^{ \mu  e *}_{LEH^5}\right)\label{40}\\
C^{e \mu u_n u_n}_{S,RR}& =& -C_{LEQU}^{ e \mu  nn} -  \frac{m_{u_n} v}{m_h^2}  C^{ e \mu }_{EH}-\frac{v^2}{\LNP^2}\left(C^{e\mu n n}_{LEQUH^2(1)}+C^{e\mu n n}_{LEQUH^2(2)}+2\frac{m_{u_n} v}{m_h^2}  C^{e \mu  }_{LEH^5}\right)\\
C^{e \mu d_n d_n}_{S,RR}& =& -  \frac{m_{d_n} v}{m_h^2}  C^{ e \mu  }_{EH}+\frac{v^2}{\LNP^2}\left(\sum_{j}V_{jn}C^{e \mu j n}_{LEQDH^2(3)}-2\frac{m_{d_n} v}{m_h^2}  C^{ e \mu  }_{LEH^5}\right)\label{42}\\
C^{e \mu u_n u_n}_{S,LR}& =&   -  \frac{m_{u_n} v}{m_h^2}  C^{  \mu e *  }_{EH}-\frac{v^2}{\LNP^2}\left(2\frac{m_{u_n} v}{m_h^2}  C^{ \mu  e *}_{LEH^5}-C^{\mu en n*}_{LEQUH^2(5)}\right)\\
C^{e \mu d_n d_n}_{S,LR}& =& \sum_{j}V_{jn}C_{LEDQ}^{* \mu e nj}  -  \frac{m_{d_n} v}{m_h^2}  C^{  \mu e *  }_{EH}+\frac{v^2}{\LNP^2}\left[\sum_{j}V_{jn}\left(C^{\mu e n j*}_{LEQDH^2(1)}+C^{\mu e n j*}_{LEQDH^2(2)}\right)-2\frac{m_{d_n} v}{m_h^2}  C^{ \mu  e *}_{LEH^5}\right]\\
C^{e \mu u_n u_n}_{S,RL}& =&  -  \frac{m_{u_n} v}{m_h^2}  C^{  e\mu  }_{EH}-\frac{v^2}{\LNP^2}\left(2\frac{m_{u_n} v}{m_h^2}  C^{ e \mu }_{LEH^5}-C^{e\mu n n}_{LEQUH^2(5)}\right)\\
C^{e \mu d_n d_n}_{S,RL}& =& \sum_{j}V^*_{jn}C_{LEDQ}^{e \mu  nj}  -  \frac{m_{d_n} v}{m_h^2}  C^{  e\mu  }_{EH}+\frac{v^2}{\LNP^2}\left[\sum_{j}V^*_{jn}\left(C^{e\mu n j}_{LEQDH^2(1)}+C^{e\mu n j}_{LEQDH^2(2)}\right)-2\frac{m_{d_n} v}{m_h^2}  C^{ e\mu}_{LEH^5}\right]
\label{46}
\eea
\bea
C^{e \mu u_n u_n}_{T,LL}& =& -C_{T,LEQU}^{*  \mu e nn}-\frac{v^2}{\LNP^2}\left(C^{\mu e n n*}_{LEQUH^2(3)}+C^{\mu e n n*}_{LEQUH^2(4)}\right)\\
C^{e \mu u_n u_n}_{T,RR}& =& -C_{T,LEQU}^{e  \mu  n n}-\frac{v^2}{\LNP^2}\left(C^{e\mu n n}_{LEQUH^2(3)}+C^{e\mu n n}_{LEQUH^2(4)}\right)\\
C^{e \mu d_n d_n }_{T,RR} &= & \frac{v^2}{\LNP^2}\sum_{j}V_{jn}C^{e\mu jn}_{LEQDH^2(5)}  \label{48}\\
C^{e \mu d_n d_n }_{T,LL} &= & \frac{v^2}{\LNP^2}\sum_{j}V^*_{jn}C^{\mu e j n*}_{LEQDH^2(5)}
\label{Tfin}
\eea
where $u_n \in \{u,c \}$, $d_n \in \{d,s,b \}$,
and
\beq
g^u_L = 1-\frac{4}{3} s_W^2~~,~~
g^u_R = -\frac{4}{3} s_W^2~~,~~
g^d_L = -1+\frac{2}{3} s_W^2~~,~~
g^d_R = \frac{2}{3} s_W^2~~~.
\eeq 
As anticipated, the low energy LFV tensors involving $d-$type quarks are matched at tree level onto the SMEFT eight dimensional tensors. Dimension eight operators could also be relevant for LL, RR scalars with $d$ quarks and RL, LR scalars with $u$ quarks, as the dimension six contributions are suppressed by Yukawa couplings.


\begin{thebibliography}{222222}

\bibitem{BW}
  W.~Buchmuller and D.~Wyler,
  ``Effective Lagrangian Analysis of New Interactions and Flavor Conservation,''
  Nucl.\ Phys.\ B {\bf 268} (1986) 621.
  doi:10.1016/0550-3213(86)90262-2





\bibitem{polonais}
B.~Grzadkowski, M.~Iskrzynski, M.~Misiak and J.~Rosiek,
  ``Dimension-Six Terms in the Standard Model Lagrangian,''
  JHEP {\bf 1010} (2010) 085
  [arXiv:1008.4884 [hep-ph]].
 

\bibitem{KO}
Y.~Kuno and Y.~Okada,
  ``Muon decay and physics beyond the standard model,''
  Rev.\ Mod.\ Phys.\  {\bf 73} (2001) 151
  doi:10.1103/RevModPhys.73.151
  [hep-ph/9909265].



\bibitem{BjW}
J.~D.~Bjorken and S.~Weinberg,
``A Mechanism for Nonconservation of Muon Number,''
Phys. Rev. Lett. \textbf{38} (1977), 622
doi:10.1103/PhysRevLett.38.622


\bibitem{BZ}
S.~M.~Barr and A.~Zee,
``Electric Dipole Moment of the Electron and of the Neutron,''
Phys. Rev. Lett. \textbf{65} (1990), 21-24
[erratum: Phys. Rev. Lett. \textbf{65} (1990), 2920]
doi:10.1103/PhysRevLett.65.21



\bibitem{W}
L.~Wolfenstein,
``Parametrization of the Kobayashi-Maskawa Matrix,''
Phys. Rev. Lett. \textbf{51} (1983), 1945

\bibitem{burashouches}
  A.~J.~Buras,
  ``Weak Hamiltonian, CP violation and rare decays,''
  hep-ph/9806471.

\bibitem{Luca}
L.~Silvestrini,
``Effective Theories for Quark Flavour Physics,''
Les Houches Lect. Notes \textbf{108} (2020)
doi:10.1093/oso/9780198855743.003.0008
[arXiv:1905.00798 [hep-ph]].

\bibitem{Huber:2005ig}
T.~Huber, E.~Lunghi, M.~Misiak and D.~Wyler,
``Electromagnetic logarithms in $\bar B \to  X_s l^+ l^-$,''
Nucl. Phys. B \textbf{740} (2006), 105-137
doi:10.1016/j.nuclphysb.2006.01.037
[arXiv:hep-ph/0512066 [hep-ph]].



\bibitem{C+C}
S.~Davidson,
``Completeness and Complementarity for $\mu \to e \gamma$, $\mu \to 3e$ and $\mu \to e$ conversion,''
[arXiv:2010.00317 [hep-ph]].

\bibitem{PSI}
    A.~Crivellin, S.~Davidson, G.~M.~Pruna and A.~Signer,
  ``Renormalisation-group improved analysis of $\mu\to e$ processes in a systematic effective-field-theory approach,''
  arXiv:1702.03020 [hep-ph].

052-016-4207-5;

\bibitem{JMT}
  R.~Alonso, E.~E.~Jenkins, A.~V.~Manohar and M.~Trott,
  ``Renormalization Group Evolution of the Standard Model Dimension Six Operators III: Gauge Coupling Dependence and Phenomenology,''
  JHEP {\bf 1404} (2014) 159
  [arXiv:1312.2014 [hep-ph]].
 E.~E.~Jenkins, A.~V.~Manohar and M.~Trott,
  ``Renormalization Group Evolution of the Standard Model Dimension Six Operators II: Yukawa Dependence,''
  JHEP {\bf 1401} (2014) 035
  doi:10.1007/JHEP01(2014)035
  [arXiv:1310.4838 [hep-ph]].


\bibitem{Murphy:2020rsh}
C.~W.~Murphy,
``Dimension-8 operators in the Standard Model Eective Field Theory,''
JHEP \textbf{10} (2020), 174
doi:10.1007/JHEP10(2020)174
[arXiv:2005.00059 [hep-ph]].




\bibitem{Herrlich:1994kh}
S.~Herrlich and U.~Nierste,
``Evanescent operators, scheme dependences and double insertions,''
Nucl. Phys. B \textbf{455} (1995), 39-58
doi:10.1016/0550-3213(95)00474-7
[arXiv:hep-ph/9412375 [hep-ph]].


\bibitem{GL}
G.~F.~Giudice and O.~Lebedev,
``Higgs-dependent Yukawa couplings,''
Phys. Lett. B \textbf{665} (2008), 79-85
doi:10.1016/j.physletb.2008.05.062
[arXiv:0804.1753 [hep-ph]].



\bibitem{CMStau}
A.~M.~Sirunyan \textit{et al.} [CMS],
``Observation of the Higgs boson decay to a pair of $\tau$ leptons with the CMS detector,''
Phys. Lett. B \textbf{779} (2018), 283-316
doi:10.1016/j.physletb.2018.02.004
[arXiv:1708.00373 [hep-ex]].


M.~Aaboud \textit{et al.} [ATLAS],
``Cross-section measurements of the Higgs boson decaying into a pair of $\tau$-leptons in proton-proton collisions at $\sqrt{s}=13$ TeV with the ATLAS detector,''
Phys. Rev. D \textbf{99} (2019), 072001
doi:10.1103/PhysRevD.99.072001
[arXiv:1811.08856 [hep-ex]].


\bibitem{CMSmuon}
G.~Aad \textit{et al.} [ATLAS],
``A search for the dimuon decay of the Standard Model Higgs boson with the ATLAS detector,''
doi:10.1016/j.physletb.2020.135980
[arXiv:2007.07830 [hep-ex]].

A.~M.~Sirunyan \textit{et al.} [CMS],
``Evidence for Higgs boson decay to a pair of muons,''
[arXiv:2009.04363 [hep-ex]].





\bibitem{CMSLFVHiggs}
A.~M.~Sirunyan \textit{et al.} [CMS],
``Search for lepton flavour violating decays of the Higgs boson to $\mu\tau$ and e$\tau$ in proton-proton collisions at $\sqrt{s}=$ 13 TeV,''
JHEP \textbf{06} (2018), 001
doi:10.1007/JHEP06(2018)001
[arXiv:1712.07173 [hep-ex]].

G.~Aad \textit{et al.} [ATLAS],
``Search for lepton-flavour-violating decays of the Higgs and $Z$ bosons with the ATLAS detector,''
Eur. Phys. J. C \textbf{77} (2017) no.2, 70
doi:10.1140/epjc/s10052-017-4624-0
[arXiv:1604.07730 [hep-ex]].




\bibitem{MEG}
  A.~M.~Baldini {\it et al.} [MEG Collaboration],
  ``Search for the lepton flavour violating decay $\mu ^+ \rightarrow \mathrm {e}^+ \gamma $ with the full dataset of the MEG experiment,''
  Eur.\ Phys.\ J.\ C {\bf 76} (2016) no.8,  434
  doi:10.1140/epjc/s10052-016-4271-x
  [arXiv:1605.05081 [hep-ex]].





\bibitem{PDB}
 K.A. Olive {\it et al.}[Particle Data Group],
 Chin.\ Phys.\ C,{\bf 38}, 090001 (2014) and 2015 update. 





\bibitem{MEGII}
  A.~M.~Baldini {\it et al.} [MEG II Collaboration],
  ``The design of the MEG II experiment,''
  Eur.\ Phys.\ J.\ C {\bf 78} (2018) no.5,  380
  doi:10.1140/epjc/s10052-018-5845-6
  [arXiv:1801.04688 [physics.ins-det]].
%
%
\bibitem{Bellgardt:1987du}
 U.~Bellgardt {\it et al.} [SINDRUM Collaboration],
 ``Search for the Decay $\mu \to 3e$,''
  Nucl.\ Phys.\ B {\bf 299} (1988) 1.
  doi:10.1016/0550-3213(88)90462-2
    
  
  
\bibitem{Mu3e}
  A.~Blondel {\it et al.},
  ``Research Proposal for an Experiment to Search for the Decay $\mu \to eee$,''
  arXiv:1301.6113 [physics.ins-det].
  

  
\bibitem{Bertl:2006up}
  W.~H.~Bertl {\it et al.} [SINDRUM II Collaboration],
  ``A Search for muon to electron conversion in muonic gold,''
  Eur.\ Phys.\ J.\ C {\bf 47} (2006) 337.
  doi:10.1140/epjc/s2006-02582-x
C.~Dohmen {\it et al.} [SINDRUM II Collaboration],
  ``Test of lepton flavor conservation in $\mu \to e$ conversion on titanium,''
  Phys.\ Lett.\ B {\bf 317} (1993) 631.
 W.~Honecker {\it et al.} [SINDRUM II Collaboration],
  ``Improved limit on the branching ratio $\mu \to e$ conversion on lead,''
  Phys.\ Rev.\ Lett.\  {\bf 76} (1996) 200.
  doi:10.1103/PhysRevLett.76.200


 
  

\bibitem{Mu2e}
 R.~M.~Carey {\it et al.} [Mu2e Collaboration],
  ``Proposal to search for $\mu^- N \to e^- N$ with a single event sensitivity below $10^{-16}$,''
  FERMILAB-PROPOSAL-0973.

\bibitem{COMET}
 Y.~G.~Cui {\it et al.} [COMET Collaboration],
  ``Conceptual design report for experimental search for lepton flavor violating mu- - e- conversion at sensitivity of 10**(-16) with a slow-extracted bunched proton beam (COMET),''
  KEK-2009-10.
M.~L.~Wong [COMET Collaboration],
  ``Overview of the COMET Phase-I experiment,''
  PoS FPCP {\bf 2015} (2015) 059.

\bibitem{tau1}
B.~Aubert \textit{et al.} [BaBar],
Phys. Rev. Lett. \textbf{104} (2010), 021802
doi:10.1103/PhysRevLett.104.021802
[arXiv:0908.2381 [hep-ex]]

\bibitem{tau2}
K.~Hayasaka, K.~Inami, Y.~Miyazaki, K.~Arinstein, V.~Aulchenko, T.~Aushev, A.~M.~Bakich, A.~Bay, K.~Belous and V.~Bhardwaj, \textit{et al.}
Phys. Lett. B \textbf{687} (2010), 139-143
doi:10.1016/j.physletb.2010.03.037
[arXiv:1001.3221 [hep-ex]].

\bibitem{belle2t3l}
E.~Kou \textit{et al.} [Belle-II],
``The Belle II Physics Book,''
PTEP \textbf{2019} (2019) no.12, 123C01
[erratum: PTEP \textbf{2020} (2020) no.2, 029201]
doi:10.1093/ptep/ptz106
[arXiv:1808.10567 [hep-ex]].



\bibitem{tau3}
Y.~Miyazaki \textit{et al.} [Belle],
Phys. Lett. B \textbf{648} (2007), 341-350
doi:10.1016/j.physletb.2007.03.027
[arXiv:hep-ex/0703009 [hep-ex]].

\bibitem{tau4}
Y.~Miyazaki \textit{et al.} [Belle],
Phys. Lett. B \textbf{699} (2011), 251-257
doi:10.1016/j.physletb.2011.04.011
[arXiv:1101.0755 [hep-ex]].

\bibitem{AS}
S.~Davidson and A.~Saporta,
``Constraints on $2\ell 2q$ operators from $\mu - e$ flavour-changing meson decays,''
Phys. Rev. D \textbf{99} (2019) no.1, 015032
doi:10.1103/PhysRevD.99.015032
[arXiv:1807.10288 [hep-ph]].



\bibitem{PP}
Y.~ Kuno {\it et al.} (PRISM collaboration), ''An Experimental Search for a 
$\mu N\to e N$ Conversion at Sensitivity of the Order of
$10^{-18}$  with a Highly Intense Muon Source: PRISM'', unpublished, J-PARC LOI, 2006.
  

\bibitem{CD}
M.~Carpentier and S.~Davidson,
``Constraints on two-lepton, two quark operators,''
Eur. Phys. J. C \textbf{70} (2010), 1071-1090
doi:10.1140/epjc/s10052-010-1482-4
[arXiv:1008.0280 [hep-ph]].

\bibitem{LRSXYZ}
H.~L.~Li, Z.~Ren, J.~Shu, M.~L.~Xiao, J.~H.~Yu and Y.~H.~Zheng,
``Complete Set of Dimension-8 Operators in the Standard Model Effective Field Theory,''
[arXiv:2005.00008 [hep-ph]].


\bibitem{NSI}
Y.~Farzan and M.~Tortola,
``Neutrino oscillations and Non-Standard Interactions,''
Front. in Phys. \textbf{6} (2018), 10
doi:10.3389/fphy.2018.00010
[arXiv:1710.09360 [hep-ph]].

\bibitem{DKUY}
S.~Davidson, Y.~Kuno, Y.~Uesaka and M.~Yamanaka,
``Probing $\mu e \gamma \gamma$ contact interactions with $\mu \to e$ conversion,''
[arXiv:2007.09612 [hep-ph]].


\bibitem{NSI3}
S.~Davidson and M.~Gorbahn,
``Charged lepton flavor change and nonstandard neutrino interactions,''
Phys. Rev. D \textbf{101} (2020) no.1, 015010
doi:10.1103/PhysRevD.101.015010
[arXiv:1909.07406 [hep-ph]].

\bibitem{luca}
  M.~Ciuchini, E.~Franco, L.~Reina and L.~Silvestrini,
  ``Leading order QCD corrections to b ---> s gamma and b ---> s g decays in three regularization schemes,''
  Nucl.\ Phys.\ B {\bf 421} (1994) 41
  doi:10.1016/0550-3213(94)90223-2
  [hep-ph/9311357].
\bibitem{czar}
A.~Czarnecki, W.~J.~Marciano and A.~Vainshtein,
  ``Refinements in electroweak contributions to the muon anomalous magnetic moment,''
  Phys.\ Rev.\ D {\bf 67} (2003) 073006
   Erratum: [Phys.\ Rev.\ D {\bf 73} (2006) 119901]
  [hep-ph/0212229].
  

\bibitem{SJL}
L.~V.~Silva, S.~J\"ager and K.~Leslie,
``Using dipole processes to constrain the flavour of four-fermion effective interactions,''
[arXiv:2012.05630 [hep-ph]].

\bibitem{BC}
G.~Buchalla and O.~Cata,
``Effective Theory of a Dynamically Broken Electroweak Standard Model at NLO,''
JHEP \textbf{07} (2012), 101
doi:10.1007/JHEP07(2012)101
[arXiv:1203.6510 [hep-ph]].

G.~Buchalla, O.~Cat\`a and C.~Krause,
``Complete Electroweak Chiral Lagrangian with a Light Higgs at NLO,''
Nucl. Phys. B \textbf{880} (2014), 552-573
[erratum: Nucl. Phys. B \textbf{913} (2016), 475-478]
doi:10.1016/j.nuclphysb.2014.01.018
[arXiv:1307.5017 [hep-ph]].

  \bibitem{P}
A.~Pich, I.~Rosell, J.~Santos and J.~J.~Sanz-Cillero,
``Low-energy signals of strongly-coupled electroweak symmetry-breaking scenarios,''
Phys. Rev. D \textbf{93} (2016) no.5, 055041
doi:10.1103/PhysRevD.93.055041
[arXiv:1510.03114 [hep-ph]].


A.~Pich, I.~Rosell, J.~Santos and J.~J.~Sanz-Cillero,
``Fingerprints of heavy scales in electroweak effective Lagrangians,''
JHEP \textbf{04} (2017), 012
doi:10.1007/JHEP04(2017)012
[arXiv:1609.06659 [hep-ph]].

\bibitem{megmW}
  S.~Davidson,
  ``Mu to e gamma and matching at mW,''
Eur.\ Phys.\ J.\ C {\bf 76} (2016) no.7,  370
  doi:10.1140/epjc/s10052-016-4207-5
  [arXiv:1601.07166 [hep-ph]].




\bibitem{DS}
W.~Dekens and P.~Stoffer,
``Low-energy effective field theory below the electroweak scale: matching at one loop,''
JHEP \textbf{10} (2019), 197
doi:10.1007/JHEP10(2019)197
[arXiv:1908.05295 [hep-ph]].



\end{thebibliography}
\end{document}